\documentclass[lettersize,journal]{IEEEtran}
\usepackage{amsmath,mathrsfs,amsfonts,amssymb}
\usepackage{algorithmic,algorithm,bm}
\usepackage{array,cuted}
\usepackage[caption=false,font=normalsize,labelfont=sf,textfont=sf]{subfig}
\usepackage{textcomp}
\usepackage{stfloats}
\usepackage{url}
\usepackage{verbatim,multirow}
\usepackage{graphicx,stfloats}
\usepackage{cite}
\usepackage{amsthm,makecell}
\begin{document}

\title{Underwater Sound Speed Profile Construction: A Review}

\author{Wei Huang~\IEEEmembership{Member,~IEEE,}, Jixuan Zhou, Fan Gao, Jiajun Lu, Sijia Li, Pengfei Wu, Junting Wang, Hao Zhang*~\IEEEmembership{Senior Member,~IEEE,}, and Tianhe Xu* 
	\thanks{Manuscript received October 15, 2023; revised XX XX, 2023.}
	\thanks{Wei Huang, Jiajun Lu, Sijia Li and Hao Zhang are with the Faculty of Information Science and Engineering, Ocean University of China, Qingdao, China (e--mail: hw@ouc.edu.cn, zhanghao@ouc.edu.cn); Tianhe Xu, Fan Gao and Junting Wang are with the Institute of Space Sciences, Shandong University, Weihai, China (e--mail:thxu@sdu.edu.cn, gaofan@sdu.edu.cn); (Corresponding author: Hao Zhang and Tianhe Xu.)}}

\markboth{IEEE Internet of Things Journal, VOL. X, NO. X, XX, 2023}%
{Huang \MakeLowercase{\textit{et al.}}: Underwater SSP construction: A Review}

\maketitle

\begin{abstract}
Real--time and accurate construction of regional sound speed profiles (SSP) is important for building underwater positioning, navigation, and timing (PNT) systems as it greatly affect the signal propagation modes such as trajectory. In this paper, we summarizes and analyzes the current research status in the field of underwater SSP construction, and the mainstream methods include direct SSP measurement and SSP inversion. In the direct measurement method, we compare the performance of popular international commercial temperature, conductivity, and depth profilers. While for the inversion methods, the framework and basic principles of matched field processing (MFP), compressive sensing (CS), and deep learning (DL) neural networks for constructing SSP are introduced, and their advantages and disadvantages are compared. The traditional direct measurement method has good accuracy performance, but it usually takes a long time, thus lacked flexibility. The proposal of SSP inversion method greatly improves the convenience and real--time performance, but the accuracy is not as good as the direct measurement method. With the continuous promotion of MFP, CS, and DL methods, the accuracy and real-time performance of sound velocity field construction algorithms have been further improved. However, the SSP inversion relies on sonar observation data, making it difficult to apply to areas that couldn't be covered by underwater observation systems, and these methods are unable to predict the distribution of sound velocity at future times. Currently, the construction method is continuously developing towards more intelligence, real--time, and precision. How to comprehensively utilize multi-source data and provide elastic sound velocity distribution estimation services with different accuracy and real-time requirements for underwater users without sonar observation data is the mainstream trend in future research on SSP construction.
\end{abstract}


\section{Introduction}
\IEEEPARstart{T}{he} construction of underwater integrated positioning, navigation, timing and communication (PNTC) systems is an important part for achieving national information construction, which is of great significance for early warning of marine disasters, underwater rescues, exploration of marine resources, and the national security \cite{Lin2020Overview}. With the development of sensing technology and underwater robotics, establishing multi-source, real-time, and three-dimensional observation networks becomes an important way for achieving underwater information perception, target identification and tracking. These networks consist of devices such as surface buoys, dive beacons, bottom anchors, autonomous underwater vehicles (AUVs), and underwater remotely operated vehicles (ROVs). 

\indent Positioning and communication are the most two important technical foundations of underwater PNTC service systems. Not only the underwater frogman, submarine and other user equipment need accurate positioning, navigation, timing services, the ocean temperature, salinity information collection and biological activity information perception also need to be accompanied by accurate positioning information, so as to demonstrate greater application value. At the same time, the perceived and collected information needs to be transmitted back to the ground information processing center in time to be utilized, so how to improve the efficiency and accuracy performance of underwater communication and positioning has been widely concerned by scholars for a long time.

\indent In the underwater environment, the attenuation of acoustic signals is much smaller than that of radio signals and optical signals, thus the acoustic signal becomes the main signal carrier of underwater PNTC system. However, because of the influence of uniformly distributed underwater temperature, salinity and pressure, the propagation speed of acoustic signals has complex dynamic spatial and temporal changing characteristics, which leads to the bending of the signal propagation path due to the Snell effect. The signal bending phenomenon poses great challenges for the efficient utilization of beam energy in directional communication systems and high-precision ranging in acoustic localization \cite{jensen2011computational}. If the regional sound speed distribution could be obtained, the signal propagation trajectory can be tracked and reconstructed according to the ray-tracing theory \cite{munk1979ocean,munk1983ocean,Munk1995Ocean}, which promotes the development of high precision of positioning, ranging, and timing as well as high efficiency of communication in the PNTC service system. Therefore, real-time and accurate sound speed field construction is of significant importance.

\indent To give a comprehensive understanding on the development of underwater sound speed field construction, we investigate and compare methods for constructing sound speed field. The contributiong of our work is summarized as:
\begin{itemize}
	\item We have done a survey on sound speed field construction that it can be obtained  by two kinds of methods: direct measurement method and inversion method.
	\item We have summarized the recognized marinstream underwater sound speed empirical formulas and compared there application scenarios.
	\item We have conducted research on advanced global commertial temperature, conductivity, and depth profiler (CTD), and compared their performance indicators.
	\item We have summarized three frameworks for constructing sound speed fields: matched field processing (MFP), compressive sensing (CS), and deep learning (DL). We have also compared the performance of different sound field construction methods.
	\item We have reviewed the development of constructing sound speed fields, identified current problems to be solved and the trends of future development.
\end{itemize}

The organization of this paper is elaborated below. Section 2 provide a classification of methods for constructing sound speed fields, summarized empirical formulas for sound speed, and compared the advanced commercial equipment used in direct measurement methods around the world. Section 3 summarize three mainstream frameworks for sound speed field inversion and provide a detailed introduction to different methods. Comparison of methods for constructing underwater sound speed field is given in Section 4, and conclusions are drawn in Section 5.

\section{Methods for Constructing Underwater Acoustic Velocity Fields}
\indent Underwater sound speed distribution has dynamic spatial and temporal difference. For a same scale of distance, the change of sound speed in the vertical direction is much larger than that of the horizontal direction, so sound speed profiles (SSPs) are usually adopted to describe the distribution of the sound speed field in a small--scale region. Currently, the mainstream methods for obtaining the underwater SSP include direct measurement and SSP inversion.
\subsection{Direct measurement methods}
\indent The attempt to measure sound velocity can be traced back to the Lake Geneva experiment in 1827. Colladon and Sturm collaborated to use sound tubes for receiving the ringing sound of a distant diving bell \cite{lasky1977review}. The Lake Geneva experiment measured the speed of sound propagation for the first time in lake water, but it was not possible to obtain the variation of sound speed with depth because of the limited experimental conditions.

\indent With the development of electronic instruments, sound velocity profiler (SVP) appeared in the 1950s, which utilizes the acoustic pulse signal for direct measurement of sound speed. Typical SVP includes sound speed change recorder developed by Brown based on the phase method in 1953, "ring-tone" SVP developed by Greenspan and Tschiegg based on the pulse method in 1957, and low-frequency SVP developed by Scheffel based on the resonance method in 1960 \cite{Nobody1977Overview}. SVP can measure the absolute value of sound speed, but the equipment is not easy to miniaturize. Moreover, SVP is not conducive to exploring the spatial-temporal changes of global sound speed distribution, so it is difficult to carry out the sound speed forecasting.

\indent In 1960, Wilson et al. summarized an accurate empirical formula of sound speed distribution \cite{Wilson1960Equation,Wang2013Marine}, which promoted the rapid development of underwater sensing equipment. 
From then on, underwater SVPs can be measured using instruments such as the CTD or more portable and smaller expendable CTD (XCTD) that works together with empirical sound speed formulas \cite{Zhang2017Review}. In addition, Leroy, Grosso, Medwin, etc. have each proposed empirical formulas under different applicable conditions, and the most commonly used empirical formulas for calculating sound speed and their scope of application are listed as follows:

\indent 1) Wilson empirical formula \cite{Wilson1960Equation}:

\begin{equation}\label{eq01}
\begin{split}
		v &= 1449.14 + v_t + v_s + v_{tpS}\\
		v_t &= 4.5721t - 4.4532 \times 10^{-2}t^2 - 2.604\times10^{-4}t^3\\
		    & \quad + 7.9851\times10^{-6}t^4\\
		v_p &= 1.60272\times10^{-1}p + 1.0268\times10^{-5}p^2\\
			& \quad + 3.5216 \times 10^{-9}p^3 - 3.3603 \times 10^{-12}p^4\\
		v_s &= 1.39799(S-35) + 1.69202\times10_{-3}(S-35)^2\\
		v_{tpS} &= (S-35)(-1.1244\times10^{-2}t+7.7711\times10^{-7}t^2\\
			& \quad +7.7016\times10^{-5}p-1.2943\times10^{-7}p^2\\
			& \quad +3.1580\times10^{-8}pt+1.5790\times10^{-9}pt^2)\\
			& \quad + p(-1.8607\times10^{-4}t + 7.4812 \times 10^{-6}t^2\\
			& \quad +4.5283\times10^{-8}t^3) + p^2(-2.5294\times10^{-7}t\\
			& \quad +1.8563\times10^{-9}t^2) + p^3(-1.9646\times10^{-10}t)
\end{split}
\end{equation}
where $t$ is the temperature in $^\circ C$, $S$ is the salinity in \textperthousand, and $p$ is the pressure in $kg/cm^3$. W.D. Wilson formula can be applied over the range of temperature $[-4^\circ C, 30^\circ C]$, salinity $[0, 37 \textperthousand]$, and pressure $[0, 1000 kg/cm^3]$.

\indent 2) Leroy empirical formula \cite{Leroy1969Equation}:

\begin{equation}\label{eq02}
	\begin{split}
		v &= 1.492.9+3(t-10)-0.006(t-10)^2 -0.04(t-18)^2\\
		& \quad +1.2(S-10)-0.01(S-35)(T-18)+Z/61\\
	\end{split}
\end{equation}
where $t$ is the temperature in $^\circ C$, $S$ is the salinity in \textperthousand, and Z is the depth in $m$. Leroy empirical formula can be applied over the range of temperature $[-2^\circ C, 34^\circ C]$, salinity$[20\textperthousand, 42\textperthousand]$, and depth up to 8000 $m$.

\indent 3) Medwin empirical formula: \cite{Medwin1975Equation}
In 1975, Medwin proposed a simplified formula based on Wilson empirical formula:

\begin{equation}\label{eq04}
	\begin{split}
		v &= 1449.2 + 4.6t - 0.055t^2 + 0.00029t^3 \\
		& \quad + (1.34-0.01t)(S-35)+0.016Z\\
	\end{split}
\end{equation}
where $t,S, and Z$ has the same meaning and unit with Leroy empirical formula. Medwin empirical formula can be applied over the range of temperature $[0, 35^\circ C]$, salinity$[0, 45\textperthousand]$, and depth up to 1000 $m$.

\indent 4) Del Grosso empirical formula \cite{Grosso1974Equation}:
In 1974, Del Grosso summarized the empirical formula for calculating sound speed under conditions with high salinity:

\begin{equation}\label{eq03}
	\begin{split}
		v &= 1402.392 + \Delta v_t + \Delta v_s + \Delta v_p + \Delta v_{tpS}\\
		\Delta v_t &= 0.501109398873\times 10t-0.550946843172\times10^{-1}t^2\\
					& \quad + 0.221535969240 \times 10^{-3}t^3\\
		\Delta v_s &= 0.132952290781\times 10S \\
					& \quad + 0.128955756844\times10^{-3}S^2\\
		\Delta v_p &= 0.156059257041\times p+0.244998688441\times 10^{-4}p^2\\
					& - 0.883392332513\times 10^{-8}p^3\\
		\Delta v_{tpS}&= -0.127562783426 \times 10^{-1}tS \\
					& \quad + 0.635191613389 \times 10^{-2}tp \\
					& \quad + 0.265484716608 \times 10^{-7}t^2p^2 \\
					& \quad - 0.159349479045 \times 10^{-5}tp^2 \\
					& \quad + 0.522116437235 \times 10^{-9}tp^3 \\
					& \quad - 0.438031096213 \times 10^{-6}t^3p \\
					& \quad - 0.161674495909 \times 10^{-8}S^2p^2 \\
					& \quad + 0.968403156410 \times 10^{-4}t^2S \\
					& \quad + 0.485639620015 \times 10^{-5}tS^2p \\
					& \quad - 0.340597039004 \times 10^{-3}tpS \\
	\end{split}
\end{equation}
where $t,S$, and $p$ has the same meaning and unit with Wilson empirical formula. Del Grosso empirical formula is applicable to the following ranges: temperature $[0, 30^\circ C]$, salinity $[30\textperthousand, 40\textperthousand]$, and pressure $[0, 1000 kg/cm^3]$.

\indent 5) Chen-Millero empirical formula \cite{Chen1977Equation}:
In 1977, Chen and Millero transmitting Wilson's data to more accurate measured sound speed data, proposed the Chen-Millero empirical formula by studying Wilson's measurements data:

\begin{equation}\label{eq05}
	\begin{split}
		v &= C_w(t,p) + A(t,p)S + B(t,p)S^{3/2} + D(t,p)S^2\\
		C_w(t,p) &= (C_{00}+C_{01}t+C_{02}t^2+C_{03}t^3+C_{04}t^4+C_{05}t^5) \\
		 & \quad +(C_{10}+C_{11}t+C_{12}t^2+C_{13}t^3+C_{14}t^4)p\\
		 & \quad +(C_{20}+C_{21}t+C_{22}t^2+C_{23}t^3+C_{24}t^4)p^2\\
		 & \quad +(C_{30}+C_{31}t+C_{32}t^2)p^3\\
		 A(t,p) &= (A_{00}+A_{01}t+A_{02}t^2+A_{03}t^3+A_{04}t^4)\\
		 & \quad +(A_{10}+A_{11}t+A_{12}t^2+A_{13}t^3+A_{14}t^4)p\\
		 & \quad +(A_{20}+A_{21}t+A_{22}t^2+A_{23}t^3)p^2\\
		 & \quad +(A_{30}+A_{31}t+A_{32}t^2)p^3\\
		 B(t,p) &= B_{00}+B_{01}t+(B_{10}+B_{11}t)p\\ 
		 D(t,p) &= D_{00}+D_{10}p\\
	\end{split}
\end{equation}
where $t,S$ has the same meaning and unit with Medwin empirical formula, $p$ is the pressure in $bar$. $C_w(t,p)$ represents the sound speed value in purified water environment. The parameters $A(t,p), B(t,p), C_w(t,p), and D(t,p)$ can be obtained by checking the table in literature \cite{Huang2016WOA}. Chen-Millero empirical formula can be applied with the following ranges: temperature $[0, 40^\circ C]$, salinity$[5\textperthousand, 40\textperthousand]$, and pressure up to 1000 $bar$. In 1994, In 1994, Millero and Li made a more accurate correction within the low-temperature range based on the Chen-Millero empirical formula and proposed the Chen-Millero-Li formula \cite{Frank1994equation}, which has been recommended by the United Nations Educational, Scientific and Cultural Organization (UNESCO) as the international standard formula for calculating underwater sound speed.

\indent 6) Coppens empirical formula \cite{Coppens1975Equation}: In 1981, Coppens extrapolated the Del Grosso's formula in the ranges of high salt, low salt, and low temperature, and then proposed the Coppens empirical formula:
\begin{equation}\label{eq06}
	\begin{split}
		v &= v_0 + (16.23+0.0253t)×0.001Z\\
		& \quad +(0.213-0.01t)×0.000001Z^2\\
		& \quad +[0.016+0.0002×(S-35.0)]×(S-35.0)×0.00001t×Z\\
		v_0 &= 1449.05+4.57×t-0.0521×t^2+0.00023×t^3\\
		& \quad +(1.333-0.0126×t+0.00009×t^2 )×(S-35.0) 
	\end{split}
\end{equation}
where $t$ is the temperature in $^\circ C$, $S$ is the salinity in \textperthousand, and Z is the depth in $m$. Coppens empirical formula can be applied within the range of: temperature $[0, 35^\circ C]$, salinity$[0, 45\textperthousand]$, and depth up to 4000 $m$. The application condition of mainstream sound speed empirical formula is given in Table~\ref{table1}

\begin{table*}[!htbp]
	\caption{comparizon of different sound speed empirical formulas\label{table1}}
	\centering
	\begin{tabular}{|c|c|c|c|c|}
		\hline
		\multirow{2}*{Empirical formula} & \multirow{2}*{Proposed year} & \multicolumn{3}{c|}{Application range}  \\
		\cline{3-5}
		\multirow{2}*{} & \multirow{2}*{} & temperature & salinity & depth/pressure \\
		\hline
		Wilson empirical formula \cite{Wilson1960Equation} & 1960 & $[-4^\circ C, 30^\circ C]$ & $[0, 37 \textperthousand]$ & $[0, 1000 kg/cm^3]$ \\
		\hline
		Leroy empirical formula \cite{Leroy1969Equation} & 1969 &$[-2^\circ C, 34^\circ C]$  &$[20\textperthousand, 42\textperthousand]$  & $[0,8000m]$ \\
		\hline
		Del Grosso empirical formula \cite{Grosso1974Equation} & 1974 & $[0, 30^\circ C]$ & $[30\textperthousand, 40\textperthousand]$ & $[0, 1000 kg/cm^3]$ \\
		\hline
		Medwin empirical formula \cite{Medwin1975Equation}& 1975 & $[0, 35^\circ C]$ & $[0, 45\textperthousand]$ & $[0,1000m]$ \\
		\hline
		Chen--Millero empirical formula \cite{Chen1977Equation} & 1977 & $[0, 40^\circ C]$ & $[5\textperthousand, 40\textperthousand]$ & $[0,1000bar]$ \\
		\hline
		Coppens empirical formula \cite{Coppens1975Equation} & 1981 & $[0, 35^\circ C]$ & $[0, 45\textperthousand]$ &  $[0,4000m]$ \\
		\hline
	\end{tabular}
\end{table*}
\indent Among the direct measurement methods of sound speed, SVP can measure the absolute value of sound speed, so it is usually used as the standard of regional sound speed distribution. However the sound speed measurement based on CTD and XCTD is more flexible and portable, but the accuracy is not as good as SVP, and different empirical formulas have different applicable ranges.

\indent Based on multi--beam bathymetry systems, Zhou et~al. studied the applicable range of mainstream sound speed empirical formulas from the perspective of determining the optimal sound speed formula \cite{Zhou2001Determination}. Leveraging SVP measurement data from the South China Sea, Chen et~al. compared and analyzed the applicable range of mainstream sound speed empirical formulas based on CTD sea trial data \cite{Chen2014Research}. According to Chen's research conclusion: the Chen-Millero-Li formula is the most suitable one for using in rivers, lakers, estuaries, coasts, and continental shelf waters. Del Grosso empirical formula and Coppens empirical formula are superior in the deep sea. Wilson empirical formula is widely mentioned in educational books because it is the earliest proposed and has better stability, but its extrapolation applicability is insufficient at the area within low-salt range.

\indent Benefiting from the advantages of simple operation, flexibility, convenience, and easy integration of miniaturized sensors, CTD has developed rapidly since the 1950s. Developed countries such as the United States, Italy, Canada, Germany, and Japan started early in the development of CTD equipment, and after decades of development, their technology has long been leading the world. Well-known commercial CTD products mainly include the Sea Bird (SBE) series CTD produced by Sea Bird company from the U.S. \cite{SBE2023}, the SonTek CastAway CTD produced by YSI from the U.S. \cite{SonTek2023}, the Ocean Seven (OS) 300 series CTD produced by Idronaut company from Italy \cite{OceanSeven2023}, RBR series CTD produced by RBR company from Canada \cite{RBR2023}, and the CTD detectors produced by Sea \& Sun Technology from Germany \cite{SeaSun2023}. While Japan pays more attention on the development of small portable product, such as XCTD products produced by tsurumi-seiki (TSK) \cite{XCTD2023}.

\begin{figure}[!t]
	\centering
	\subfloat[]{\includegraphics[width=2.5in]{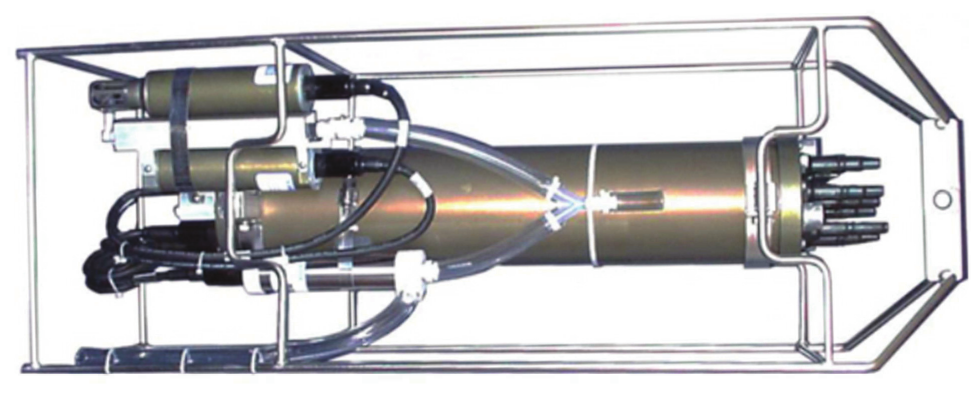}%
		\label{fig01a}}
	\hfil
	\subfloat[]{\includegraphics[width=2.5in]{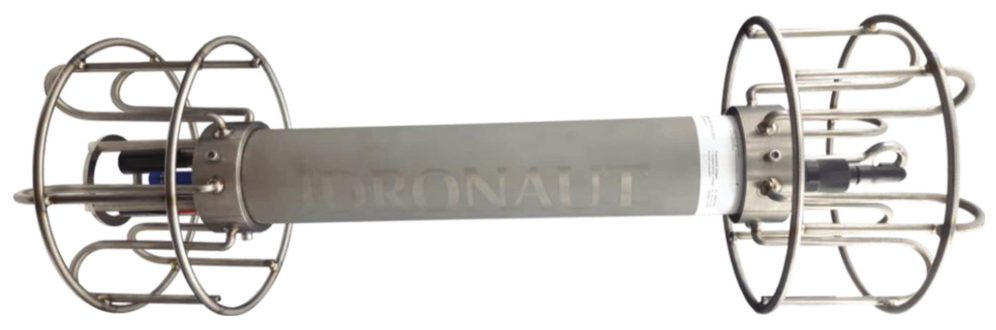}%
		\label{fig01b}}
	\hfil
	\subfloat[]{\includegraphics[width=2.5in]{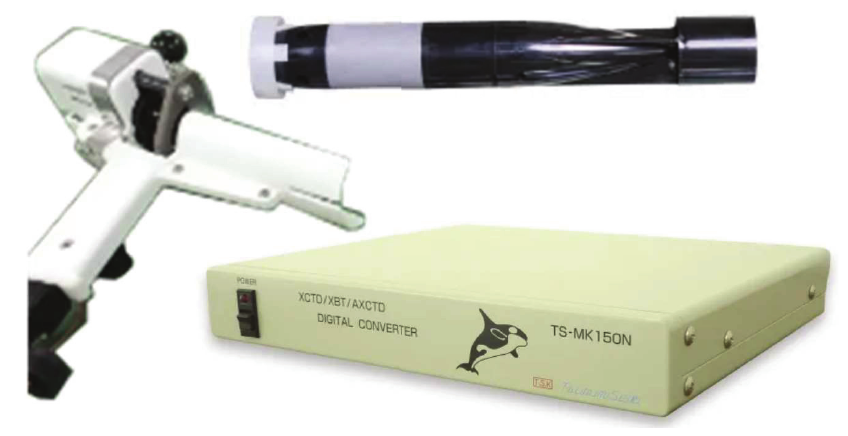}%
		\label{fig01c}}
	\hfil
	\subfloat[]{\includegraphics[width=2.5in]{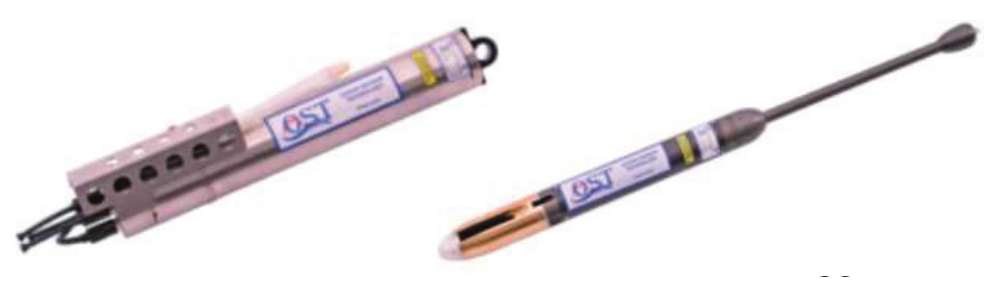}%
		\label{fig01d}}
	\caption{Typical commercial CTDs and XCTDs. (a) SBE 911 plus by Sea Bird Inc.. (b) OS 320 plus by Idronaut. (c) XCTD by TSK. (d) OST by National Ocean Technology Center of China.}
	\label{fig01}
\end{figure}

\indent The development of CTD equipment in China started relatively late, however, after years of development, many types of fully independent technical products have appear based on shipboard, fixed platforms or underwater mobile platforms. The National Ocean Technology Center is the earliest unit to carry out CTD technology research in China, and gradually formed Ocean Sensor Technology (OST) series products from 2020, which reached the international level in temperature and salinity measurement \cite{OST2023}. Although China's temperature and salinity depth measurement technology has made great breakthroughs, there is still a certain gap in the depth coverage range compared to the top commercial CTDs due to the shortcoming of pressure resistance technology. Parts of commercial CTDs are given in Fig.~\ref{fig01}. The performance comparison of typical commercial CTD/XCTD is shown in Table~{table2}.

\begin{table*}[!htbp]
	\caption{comparizon of commertial CTD and XCTD\label{table2}}
	\centering
	\begin{tabular}{|c|c|c|c|c|c|c|c|}
		\hline
		\multicolumn{2}{|c|}{Model} & \makecell{OST15D \\ \cite{OST2023}} & \makecell{SBE911plus \\\cite{SBE2023}}  & \makecell{OS320plus \\\cite{OceanSeven2023}}  & \makecell{CTD90M\\ \cite{SeaSun2023}}  & \makecell{RBR Concerto3\\ \cite{RBR2023}} & \makecell{XCTD-4N \\\cite{XCTD2023}}\\
		\hline
		\multirow{4}*{Temperature ($^\circ C$)} & Range  & [-5,35] & [-5,35] & [-5,45] & [-2,36] & [-5,35] & [-2,35]\\
		\cline{2-8}
		\multirow{4}*{} & Accuracy  & $\pm 0.002$ & $\pm 0.001$ & $\pm 0.001$ & $\pm 0.002$ & $\pm 0.002$ & $\pm 0.02$ \\
		\cline{2-8}
		\multirow{4}*{} & Resolution  & 0.0001 & 0.0002 & 0.0001 &  0.0005 & < 0.00005 &  0.01 \\
		\cline{2-8}
		\multirow{4}*{} & Stability  & $\pm 0.002$/year & $\pm 0.0002$/month & - & - & $\pm 0.002$/year & - \\
		\hline
		\multirow{4}*{Conductivity ($S/m$)} & Range  & [0,7] & [0,7] & [0,7] & [0,30] & [0,8.5] & [0,6]\\
		\cline{2-8}
		\multirow{4}*{} & Accuracy  & $\pm 0.0003$ & $\pm 0.0003$ & $\pm 0.0001$ & $\pm 0.001$ & $\pm 0.0003$ & $\pm 0.003$ \\
		\cline{2-8}
		\multirow{4}*{} & Resolution  & 0.00001 & 0.00004 & 0.00001 &  0.0005 & < 0.0001 &  0.0015 \\
		\cline{2-8}
		\multirow{4}*{} & Stability  & $\pm 0.0003$/month & $\pm 0.0003$/month & - & - & $\pm 0.001$/year & - \\
		\hline
		\multirow{4}*{Pressure ($10^4kPa$)} & Range  & [0,7] & [0,10.5] & [0,10] & [0,6] & [0,6] & [0,1.85]\\
		\cline{2-8}
		\multirow{4}*{} & Accuracy  & $\pm 0.05\%$ range & $\pm 0.015\%$ range & $\pm 0.01\%$ range & $\pm 0.05\%$ range & $\pm 0.05\%$ range & - \\
		\cline{2-8}
		\multirow{4}*{} & Resolution  & $\pm 0.001\%$ range & $\pm 0.001\%$ range & $\pm 0.002\%$ range & $\pm 0.002\%$ range & $\pm 0.001\%$ range & - \\
		\cline{2-8}
		\multirow{4}*{} & Stability  & $\pm 0.05$ range/year & $\pm 0.02$ range/year & - & - & $\pm 0.05$ range/year & - \\
		\hline
		\multicolumn{2}{|c|}{Department} & Oceans Center & Sea Bird & Idronaut  & SST  & RBR & TSK \\
		\cline{3-8}
		\multicolumn{2}{|c|}{Country} & China & the U.S. & Italy & Germany & Canada & Japan \\
		\hline
	\end{tabular}
\end{table*}

\begin{figure}[!t]
	\centering
	\subfloat[]{\includegraphics[width=2.5in]{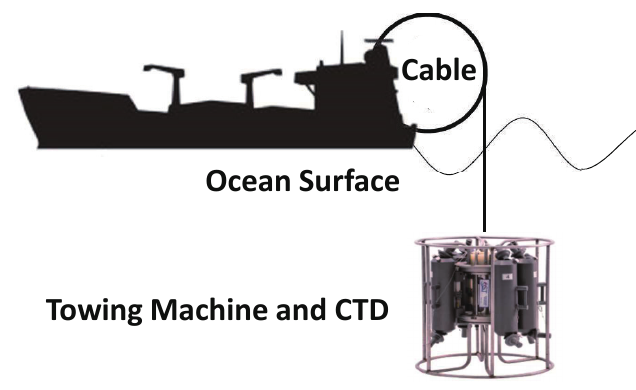}%
		\label{fig02a}}
	\hfil
	\subfloat[]{\includegraphics[width=2.5in]{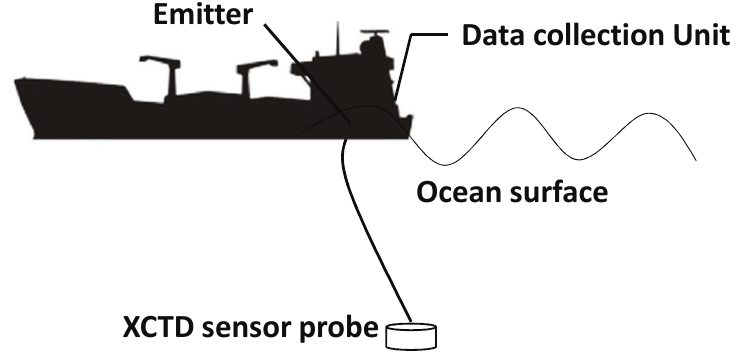}%
		\label{fig02b}}
	\caption{Ship assisted SSP measurement. (a) CTD. (b) XCTD.}
	\label{fig02}
\end{figure}

\indent The advantage of utilizing CTD (or SVP) for SSP measurement lies in good accuracy performance and high depth resolution. However, since the measurement needs to be executed by stopping the ship with the help of the shipborne towing machine as shown in Fig.~\ref{fig02a}. Thus, the measurement period is quite long, which seriously affects the operational efficiency. For example, when the down and up speed of the cable is 50 $m/min$, it takes 80 minutes to measure the SSP at a depth of 2000 meters. In contrast, XCTD measurement can be executed while the ship platform keeps sailing (within 5 knots). The probe falls by gravity, the data is transmitted back through the copper wire, and the probe is discarded at the end without recycling, so it does not affect the operational efficiency. For example, it only takes about 20 minutes to measure the SSP at a depth of 2,000 $m$. Meanwhile, the portability of XCTD allows it to be carried on automated platforms such as drones or unmanned underwater vehicles (UUVs) \cite{Zhang2017Review,Trampp2012Upper,Shi2015Expendable}, but the shortcoming of XCTD is that its depth coverage usually does not exceed 2,000 $m$ because of the limitation of the sensor's pressure resistance performance. Fig.~\ref{fig02b} shows a schematic diagram of the SSP measurement using XCTD.

\indent In order to reconstruct the full ocean depth SSP using partial ocean depth sound speed data measured by XCTD, Cheng et~al. proposed an SSP extension method for land-marginal deep-water areas based on the orthogonal empirical function (EOF) decomposition \cite{Cheng2016SSP}.  Cheng effectively realized the low-precision loss extension of the SSPs. However, it requires the prior data information coverage to exceed the depth of the sound channel axis, so the application is limited. Inspired by the above method, Huang et~al. also proposed an SSP extension method based on EOF decomposition \cite{Huang2023Fast}, which can effectively realize the extension of XCTD measurement data in the global ocean area, while doesn't need the prior data information coverage to exceed the depth of the sound channel axis. 

\indent With the increased demand for ocean mapping information, low-cost, large-scale ocean mapping operations comes into attention. In the early 21st century, the Canadian BOT company first developed the moving vessel profiler (MVP), forming the MVP30, MVP100, MVP200, MVP300 and MVP800 series products \cite{MVP3002004}. The main components of the MVP system include a towing machine, a data acquisition unit, a winch, a hydraulic system and a remote control system, etc. The towing machine can not only be equipped with CTD, but also multiple functional marine mapping equipment such as chlorophyll meters and dissolved oxygen sensors.

\indent In 2005, the First Institute of Oceanography of the State Oceanic Administration (SOA) developed a domestic MVP system for the first time in China \cite{Wang2005Oceanography}, which was verified by the experiments of 0-200 $m$ depth, indicating that the product reached the international level at that time. Since then, the domestic MVP system has been gradually updated and iterated, such as the CZT1-3A MVP system developed by the 715 Research Institute of China in 2010 \cite{Dual2010}, and the MVP 3400 system developed by the Institute of Oceanography of the Chinese Academy of Sciences in conjunction with the Shanghai Laurel Instrument Systems Co. in 2016 \cite{Ren2022MVP}. The MVP 3400 has the maximum depth coverage of 3400 $m$ with an operating cycle of 70 minutes when the ship is stopping. At a speed of 12 knots, the operating depth can reach 300 $m$, with a cycle of 13 minutes.

\subsection{Sound speed inversion method}
\indent The heterogeneous spatial and temporal distribution of sound speed leads to the dynamic variability of the underwater sound field. In 1979, Munk and Wunsch introduced the method of geophysical inversion into ocean physics, and creatively put forward the concept of ocean acoustic tomography (OAT) \cite{munk1979ocean,munk1983ocean,Munk1995Ocean}, which includes the forward evolution problem, the observation technique and the inversion problem. The forward evolution problem focuses on the acoustic propagation model, which predicts the theoretical sound field under a given topology of the source and the sink. The observation technique focuses on the accurate measurement of the acoustic field, so as to obtain the signal propagation information based on the observation data. The inversion problem estimates the ocean physical parameters through the optimization algorithms using the observation data of the acoustic field, such as the signal arrival strength and the signal propagation time.

\indent SSP inversion is a kind of ocean acoustic tomography, which inverts the distribution of sound speed in the target spatial and temporal intervals using the ocean acoustic field observation data. In 1981, Woods Hole Oceanographic Institution, jointly with Massachusetts Institute of Technology (MIT) and other organizations, carried out an ocean experiments in the test area of the Gulf Stream using sound generators, sound receivers, acoustic navigation systems, and other equipment, and estimated the regional SSPs through the acoustic tomography method, which in practice validated the feasibility of SSP inversion methods \cite{cornuelle1989ocean}. Compared with the direct measurement methods using equipment such as CTD, SVP or XCTD, the SSP inversion method adopts the pre-deployed underwater anchors to measure the acoustic field, which has a shorter response period and is more in line with the real-time demand in the fields such as underwater disaster warning, national defense and security. 

\section{Inversion Technology of Underwater SSP}
\indent In the past 30 years, scholars have carried out a lot of research work on SSP inversion methods, and have successively proposed a variety of methodological frameworks, such as MFP, CS, and DL.
\subsection{MFP for SSP Inversion}
\begin{figure}[!htbp]
	\centering
	\includegraphics[width=0.8\linewidth]{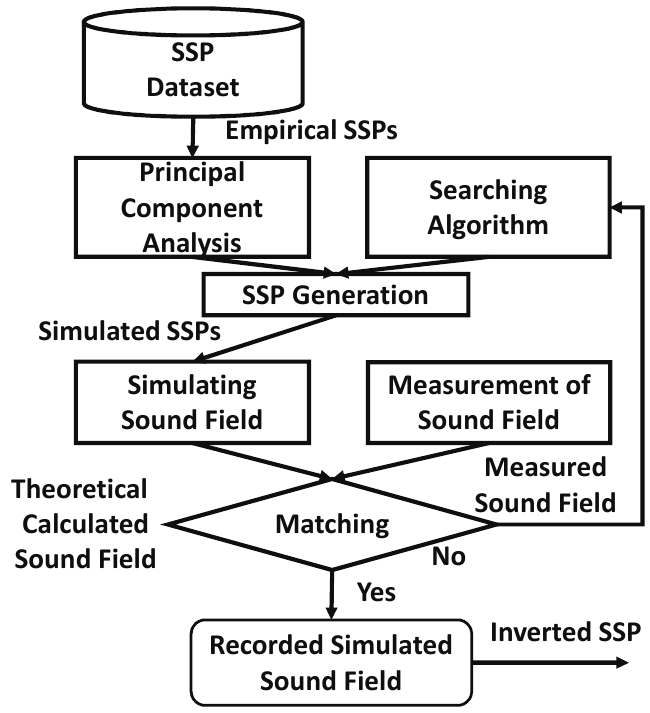}
	\caption{MFP framework for SSP Inversion.}
	\label{fig03}
\end{figure}
\indent In 1991, the U.S. Naval Laboratory applied the MFP method \cite{tolstoy1991acoustic} as shown in Fig.~\ref{fig03} to ocean acoustic tomography, which has established the mainstream framework for the inversion of SSPs for a long time. In the framework of MFP, the principal component analysis firstly is performed on the empirical SSPs to extract their principal component features in the target spatial and temporal intervals. Secondly, a searching algorithm is used to find the principal component coefficients for generating candidate SSPs. Then, for each candidate SSP, the theoretical sound field distributions are calculated according to the ray theory or the normal mode theory. Finally, the simulated and calculated sound field is matched with the actual measured values of the ocean sound field, and the sound speed distribution corresponding to the optimal sound field matching result is the final SSP inversion result. The MFP method avoids the difficulty in establishing the inverse mapping relationship from sound field information to sound speed distribution, and provides an effective way for the construction of SSPs.

\subsubsection{EOF based MFP method}
\indent In MFP, principal component analysis of the empirical SSP data is requited to obtain its distribution characteristics. The MFP method based on principal component analysis by EOF was firstly used for SSP inversion by Tolstoy in 1991 \cite{tolstoy1991acoustic}, and the lattice-point traversal method was used to search for the matching term, which has high computational complexity in the process and the time consumption of inversion needs to be reduced.

\indent EOF is a principal component analysis method that extracts the main data feature quantities by analyzing the structural features in the matrix data. For a given sound speed inversion task, a set of historical empirical SSPs is denoted as:
\begin{equation}\label{eq07}
\bm{\mathcal{S}_{p}} = \left\{\bm{\mathcal{S}_{p,1}},\bm{\mathcal{S}_{p,2}},...,\bm{\mathcal{S}_{p,i}},i=1,2,...,I\right\},
\end{equation}
where $\mathcal{S}_{p,i} = \left\{(d_i,s_{p,i,1}),...,(d_i,s_{p,i,j})\right\},j=1,2,...,J$ is the $i$th sample, $d_i$ is the depth of the $j$th layer, $s_{p,i,j}$ is the corresponding sound speed value. According to empirical SSPs, the average SSP $\mathcal{\bar{S}}_p$ can be calculated as:

\begin{equation}\label{eq08}
\bm{\bar{\mathcal{S}}_{p}}=\left\{(d_1,\bar{s}_{p,1}),(d_2,\bar{s}_{p,2}),...,(d_j,\bar{s}_{p,j})\right\},
\end{equation}
where $\bar{s}_{p,j}=\frac{1}{I}\left(s_{p,1,j} + s_{p,2,j}+ ...+ s_{p,i,j}\right)$ is the average sound speed value at the $j$th depth layer. Let $\bm{\mathcal{S}_{p,i}^{s}} = \left[s_{p,i,1},s_{p,i,2},...,s_{p,i,j}\right]$ and $\bm{\bar{\mathcal{S}}_{p}^{s}} = \left[\bar{s}_{p,1},\bar{s}_{p,2},...,\bar{s}_{p,j}\right]$ represent the correspond part of sound speed value. Then, a residual matrix can be constructed as:

\begin{equation}\label{eq09}
	\bm{\mathcal{S}}_{X,r}=\left[\bm{\mathcal{S}_{p,1}^{s}}-\bm{\bar{\mathcal{S}}_{p}^{s}},\bm{\mathcal{S}_{p,2}^{s}}-\bm{\bar{\mathcal{S}}_{p}^{s}},...,\bm{\mathcal{S}_{p,i}^{s}}-\bm{\bar{\mathcal{S}}_{p}^{s}}\right].
\end{equation}
Based on the above residual matrix, there will be a covariance matrix:

\begin{equation}\label{eq10}
	\bm{\mathcal{C}}_{S}=\frac{1}{I}\bm{\mathcal{S}}_{X,r}\times\bm{\mathcal{S}}_{X,r}^T.
\end{equation}
The covariance matrix $\bm{\mathcal{C}}_{S}$ satisfies:

\begin{equation}\label{eq11}
	\bm{\mathcal{C}}_{S}\times \bm{\mathcal{U}}=\bm{\mathcal{U}} \times \bm{\Lambda},
\end{equation}
where $\bm{\mathcal{U}}=[u_1,...,u_m],m=1,2,...,M$ is a $J\times M$--orders matrix, and each element of it is an eigenvector. $\bm{\Lambda}$ is an $M$--order diagonal matrix, with diagonal elements representing the magnitude of the eigenvalues corresponding to the eigenvectors. The magnitude of the eigenvalues indicates the importance of the eigenvectors in reconstructing the SSP distribution of the target region.

\indent For any empirical SSP $\bm{\mathcal{S}_{p,i}^{s}}$, the projection coefficient can be obtained by projecting the residual between the average SSP and itself onto the eigenvectors:

\begin{equation}\label{eq12}
	\bm{f_{i}}= \bm{\mathcal{U}}^T  \left[ \bm{\mathcal{S}_{p,i}^{s}} - \bm{\mathcal{\bar{S}}_{p}^{s}} \right],
\end{equation}
where the projection coefficient $bm{f_{i}}$ is an $M\times1$--order vector. Then, empirical SSP samples can be reconstructed by using the following equation:

\begin{equation}\label{eq13}
	\bm{\mathcal{S}_{p,i}^{s}} = \bm{\mathcal{\bar{S}}_{p}^{s}} + \bm{\mathcal{U}}\bm{f_{i}}.
\end{equation}

\indent In MFP, heuristic algorithms are adopted to search for combinations of EOF eigenvectors and coefficients, 
so as to generate different candidate SSPs $\bm{\breve \mathcal{S}_p^s}$. For each candidate SSP, the sound field distributions can be simulated through the ray theory or the normal mode theory. Finally, by optimally matching the simulated sound fields with the actual measured sound fields, the SSP inversion result is determined from the candidate SSPs.

\indent To improve the accuracy of sound speed inversion, Zhang studied the issue that MFP is easily affected by seabed parameter mismatch. In 2002, he proposed an SSP inversion method based on matching beam forming \cite{zhang2002beam}, which reduces the number of seabed reflections through beam propagation path control and to some extent reduces the dependence of inversion accuracy on seabed parameters. Zhang from Harbin Engineering University \cite{zhang2012inversion,zhang2013inversion}, studied the propagation law during reflection from uneven seabed and proposed a three-dimensional spatial feature sound line searching and propagation calculation model in 2013, which explored the effects of parameter mismatches such as seafloor depth, sound source location and array tilt on the performance of SSP inversion.

\indent In order to accelerate the searching speed of matching feature terms in MFP, Zheng et~al.  proposed an improvement algorithm based on perturbation method in 2017 \cite{zheng2017improved}, which transforms the SSP inversion from nonlinear optimization into the form of linear equations to reduce the time consumption under the condition of sacrificing the partial accuracy. Some researchers have introduced heuristic algorithms in MFP to speed up the inversion process, such as particle swarm optimization (PSO) \cite{zhang2012inversion,zhang2013inversion}, simulated annealing (SA) \cite{zhang2005thestudy}, genetic algorithm (GA) \cite{tang2006sound,sun2016inversion}. Compared with the lattice point traversal search method in the literature \cite{tolstoy1991acoustic}, the heuristic algorithm speeds up the process of searching the principal component coefficients of the eigenvectors of the SSP, but the core idea of heuristic algorithms is based on the Monte Carlo thought, and it needs to set up a sufficiently large number of particles (e.g., particle swarm optimization) or number of populations (e.g., genetic algorithm) to guarantee the search probability of the optimal or sub--optimal matching terms, and thus it still has a high amount of computational complexity.

\indent Different from the traditional active sonar communication method, Xu et~al. proposed an MFP--based SSP inversion method using AUV radiating noise as a sound source in 2016, focusing on researching and solving the problem of SSP inversion with the impact of Doppler interference\cite{zhang2015inversion}.

\subsubsection{MFP Method Based on Dictionary Learning}
\indent In 2017, Bianco et~al. from the University of California, San Diego proposed a dictionary learning based method for SSPs inversion \cite{Bianco2017Dictionary,Bianco2019Machine}, which focuses on sparse representation of SSPs. Sparse matrix is used instead of EOF method, and dictionary learning is achieved by combining singular value decomposition (SVD) and k-means, called K-SVD algorithm. The SVD algorithm sparsely expresses the empirical SSP, and any sample can be represented as a linear combination of dictionary sparse vectors added to the average SSP distribution:

\begin{equation}\label{eq14}
	\bm{\mathcal{S}_{p,i}^{s}} = \bm{\mathcal{\bar{S}}_{p}^{s}} + \bm{\mathcal{Q}}\bm{q_{i}}.
\end{equation}
where $\bm{\mathcal{Q}}$ is the sparse matrix of SSPs. If the number of sampling point for any SSP is $J$, then $\bm{\mathcal{Q}}$ will be a $K\times J$--order matrix, and there is usually $K\ll J$. $\bm{q_{i}}$ is a $K\times 1$--order vector of coefficients.

\indent The dictionary vectors constructed by dictionary learning do not need to be orthogonal to each other, which can effectively improve the efficiency of feature vectors in expressing the principal components of SSPs and reduce the number of feature vectors and coefficients. However, this method still belongs to the category of MFP, and the high searching complexity of matching terms has not been solved.

\subsection{CS for SSP Inversion}
\begin{figure}[!htbp]
	\centering
	\includegraphics[width=0.6\linewidth]{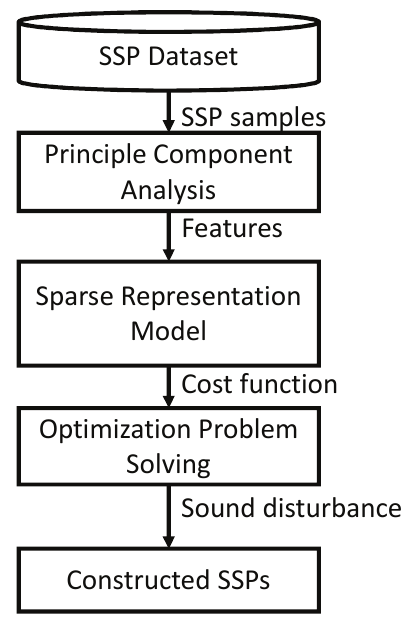}
	\caption{CS framework for SSP Inversion.}
	\label{fig04}
\end{figure}

\indent Bianco et~al. \cite{Bianco2016CS} and Choo \cite{choo2018compressive} proposed a CS--based SSP inversion method in 2016 and 2018, respectively. By using signal propagation intensity and signal propagation time, a CS dictionary was established which describes the impact of sparse sound speed disturbances on the sound field. The least squares method was used to solve the over--determined problem. The CS framework is shown in Fig.~\ref{fig04}.

\indent The dictionary constructed in the CS framework is different from that of the dictionary learning MFP. The dictionary of the latter is a sparse principal component representation of the empirical SSPs in the region. In addition to the above parts, the dictionary construction in the CS framework also establishes a mapping relationship from sound speed disturbance to sound field distribution.

\indent Let $\bm{q}$ represent a set of sparse dictionary coefficients for SSPs, and the actual measured values of sound field observations $sf_(p,t)$ can be expressed as:

\begin{equation}\label{eq15}
	\bm{sf_{p,t}} = g(\bm{q}) + \bm{ns}.
\end{equation}
where, $\bm{ns}$ is the observation noise, $g(\cdot)$ is the mapping function from SSP to sound field distribution.
For the normal mode theory, the received signal strength information can be obtained through equation \eqref{eq15}
and for the ray theory, the signal propagation time information can be obtained. 

\indent \eqref{eq15} is difficult to be directly solved because of the fact that $g(\cdot)$ is usually a nonlinear function. In the CS theory, with small disturbances of sound speed (i.e. a small q-value space), the above equation is expanded by Taylor series and the first-order Taylor expansion is saved to obtain the estimated sound field:

\begin{equation}\label{eq16}
	\overline{\bm{sf}}_{p,t} = g(\bm{0}) + \frac{\partial g(\bm{q})}{\partial \bm{q}}|_{\bm{q=0}}\bm{q}=g(\bm{0})+\bm{\mathcal{D}q}.
\end{equation}
where $\mathcal{D}$ is the dictionary matrix. Then the optimization objective function is:

\begin{equation}\label{eq16}
	\hat{\bm{q}}=\arg \min_{\bm{q}}\left\| \overline{\bm{sf}}_{p,t}-\bm{sf}_{p,t}\right\|_2^2+\mu |\bm{q}|_1,
\end{equation}
where $|\bm{q}|_1$ is the regularization term, and $\mu$ is the regularization coefficients. The objective function can be solved by CVX tools or orthogonal matching pursuit (OMP) algorithms \cite{Bianco2016CS,choo2018compressive}.

\indent In CS--based SSP inversion methods, due to the fact that the principal component coefficients can be obtained through fewer least squares iterations, the real-time inversion performance is greatly improved compared to the MFP. However, the iterative process makes the real-time performance weaker than the neural network model. In addition, the dictionary establishment process adopts first-order Taylor expansion for linear approximation, which sacrifices the inversion accuracy.

\subsection{DL for SSP inversion}
\subsubsection{Neural network inversion of SSP based on sound field observation data}
\begin{figure}[!htbp]
	\centering
	\includegraphics[width=\linewidth]{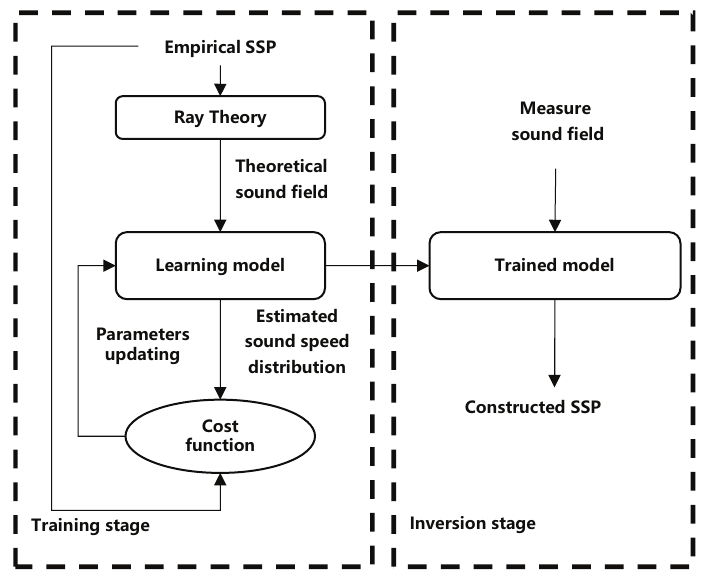}
	\caption{DL framework for SSP Inversion.}
	\label{fig05}
\end{figure}

\indent In recent years, machine learning has been widely applied in regression and classification problems in multiple fields such as image recognition and numerical prediction. Benefiting from its ability to fit complex nonlinear relationships, it is very suitable for solving the problem of inversion of sound speed distribution from sound field distribution. In 1995, Stephan, first established a framework for inversion of sound speed fields using artificial neural networks (ANNs) \cite{Stephan1995Tomographic}, as shown in Fig.~\ref{fig05}.

\indent The main idea of deep learning ANN for constructing SSPs is to invert the SSP by mining the implicit sound speed distribution information in the sound field data. For empirical SSP samples, under the condition of known topological relationships between the signal source and  receivers, sound field simulation is carried out using normal mode theory, ray theory, or parabolic equation models. The simulated sound field information is used as input data for the model and the labeled values of the SSP are used as output. During the training stage, the model's parameters are adjusted according to established cost functions, which establishes a mapping relationship from sound field to sound speed field. At the inversion stage, by putting real-time measured sound field information into the model, the SSP of the target area can be constructed.

\indent The distribution of sound speed has spatial and temporal differences, and the advantage of deep learning ANN models for constructing SSPs is that specialized models can be pre-set for different spatial and temporal intervals during model establishment. Then, the training of model can be completed offline in advance. After model convergence, for inputting sound field data, only one turn of forward iteration matrix operation is required to obtain the estimated SSP. Therefore, compared to MFP--based and CS--based SSP inversion methods, its inversion stage has significant real-time advantages. However, the ANN model also faces two important challenges when used to construct SSPs. On the one hand, underwater sound field measurement is inevitably affected by complex underwater environmental coupling interference such as noise and multipath effects, which will reduce the accuracy of the model. On the other hand, ANN model training requires a large number of empirical SSPs as reference data samples. Due to the high cost of obtaining empirical SSP data through instruments such as CTD or SVP, there are problems of insufficient training sample data in many ocean areas, which can easily lead to accuracy performance decreasing due to the overfitting phenomenon.

\begin{figure}[!htbp]
	\centering
	\includegraphics[width=\linewidth]{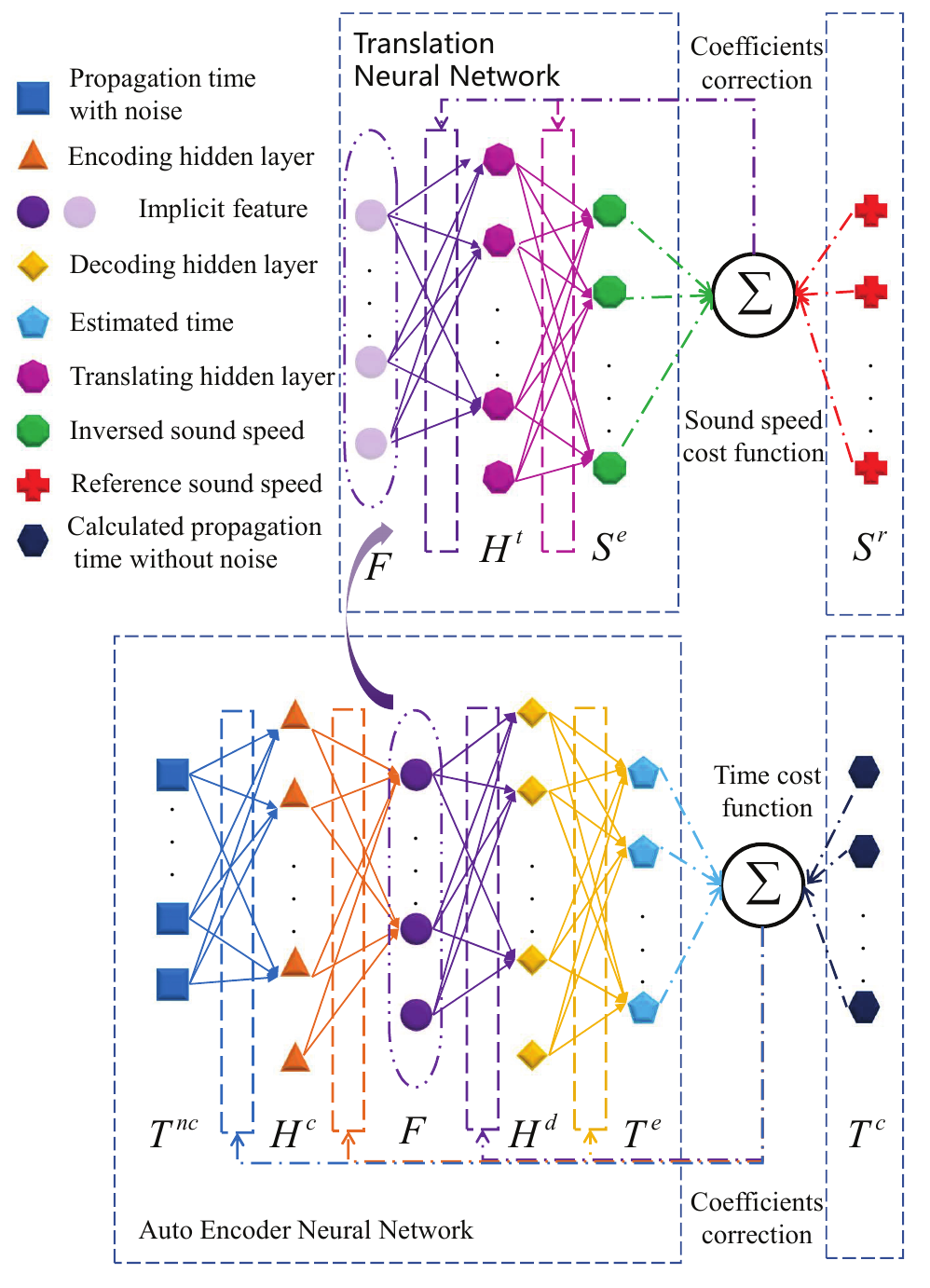}
	\caption{AEFMNN model for SSP Inversion.}
	\label{fig06}
\end{figure}

\indent In order to improve the robustness ability of DL--based SSP construction models against noise in sound field measurement, the author of this paper proposed an auto-encoding feature mapping neural network (AEFMNN) structure in 2021 \cite{Huang2019Collaborating}, as shown in Fig.~\ref{fig06}. By using an auto-encoding neural network to reconstruct noisy sound fields into denoised sound field, stronger robust hidden features are extracted. Then, the feature mapping network is used to establish a mapping relationship from hidden features to the distribution of SSPs.

\indent To address the problem of overfitting in few-shot learning scenarios, the author proposed an ANN model for SSP construction with intelligent extended dataset \cite{huang2018underwater} based on \cite{Huang2019Collaborating}. This method first extracts feature points from existing reference samples, and then generates new SSP samples based on the extracted feature points to expand the reference sample dataset (the expanded samples do not exceed $30\%$ of the total samples). This method to some extent alleviates the situation of insufficient reference sample data, but still has certain requirements for the amount of original reference sample data.

\begin{figure}[!htbp]
	\centering
	\includegraphics[width=\linewidth]{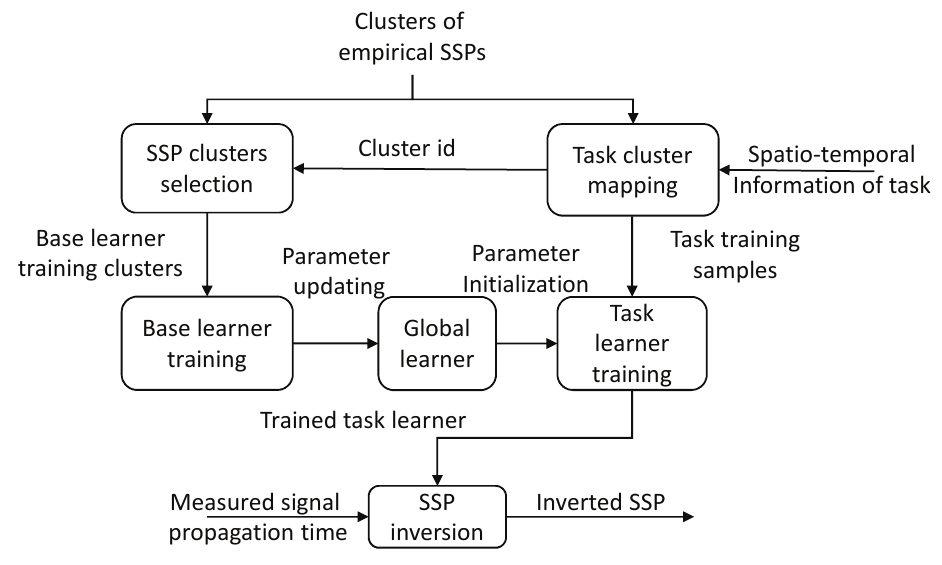}
	\caption{TDML model for SSP Inversion.}
	\label{fig07}
\end{figure}

\indent In order to effectively address the overfitting effect in few-shot learning, in 2021, references \cite{Huang2023Meta} proposed a task driven meta learning (TDML) framework for constructing SSPs. The model is shown in Fig.~\ref{fig07}. TDML includes multiple base learners, a global learner, and on-demand task learners. 
The core idea is to use multiple base learners to simultaneously learn different spatio--temporal regions and types of SSPs, obtain common features of SSP distribution, and pass them on to the task learner, enabling the task learner to achieve model convergence with a small number of samples and training. Through this process, sensitivity of the model to input data can be maintained, and thereby improving its ability to cope with overfitting.

\begin{figure}[!htbp]
	\centering
	\includegraphics[width=\linewidth]{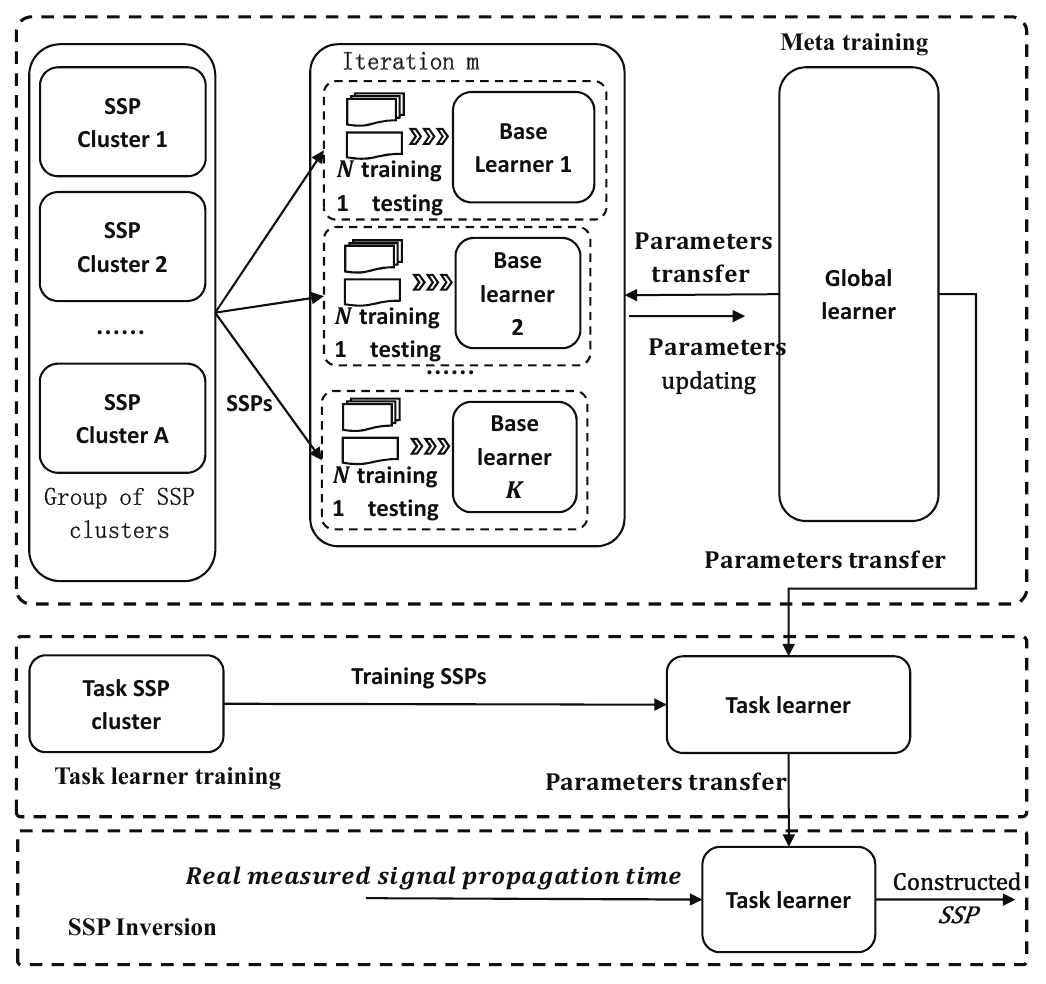}
	\caption{Training and inversion of SSP based on TDML model.}
	\label{fig08}
\end{figure}

\indent The model training method of TDML is shown in Fig.~\ref{fig08}. During the meta--model (the base learners and the global learner) training stage, for each round of parameter updating, K samples are randomly selected from a set of A SSP types and assigned to K base learners one by one. Each base learner uses N samples for training and 1 sample for testing. This training method is called K-way N-shot.

\indent The aforementioned ANN, AEFMNN, and TDML methods for SSP inversion all use neural network models to establish a mapping relationship from sound field to sound speed distribution, which can quickly estimate the distribution of SSPs based on real--time sound field input data. Howeve, due to the economic costs of device development and limited coverage range, these methods, similar to traditional MFP and CS methods, have spatial constraints on their applications.

\subsubsection{Constructing SSPs using neural networks without sound field observation data}
\indent In order to construct an SSP in a spatio--temporal interval without obtaining sound field observation data, reference \cite{Huang2019Collaborating} provides a strategy for constructing a large--scale three-dimensional sound speed field using multiple base stations in different regions. The application scenario of reference \cite{Huang2019Collaborating} is for a distributed seabed observation system, where multiple pairs of buoys or anchors with relatively concentrated spatial orientations form a base station group (which can communicate with each other internally). By measuring the sound field at each base station group, the SSP of the base station group area is inverted. After inverting the SSPs of multiple base station groups, the three-dimensional sound distribution of the whole area can be obtained through interpolating. Ultimately, the sound speed field within the closed area of the base station group can be constructed, but it does not reduce equipment development costs.

\begin{figure}[!htbp]
	\centering
	\includegraphics[width=\linewidth]{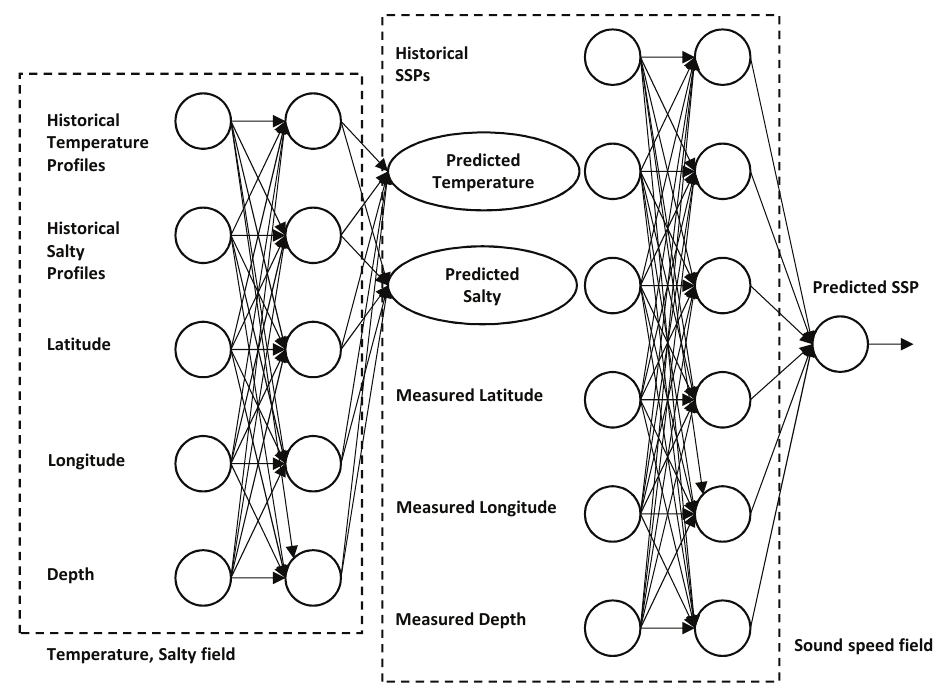}
	\caption{RBF model for SSP inversion.}
	\label{fig09}
\end{figure}

\indent In 2020, Yu et~al. \cite{Yu2020SSP} and Wang \cite{Wang2022Ocean} proposed a SSP prediction method based on radial basis function (RBF) neural network. The model structure is shown in Fig.~\ref{fig09}. This method sets up a network for constructing temperature and salinity fields, as well as a network for constructing sound speed fields. Firstly, it utilizes historical temperature and salinity profiles and their sampling longitude, latitude, and depth information to construct the average temperature and salinity fields in the target area, providing constraints for constructing a network for SSPs. Then, the average SSP of the target area is constructed using the historical SSP and its sampling longitude, latitude, and depth information. The RBF method does not require on-site measurement of sound field observation data, but the model construction results tend to approach the mean of regional samples, making it difficult to accurately describe the temporal and spatial distribution changes of SSPs.

\begin{figure}[!htbp]
	\centering
	\includegraphics[width=\linewidth]{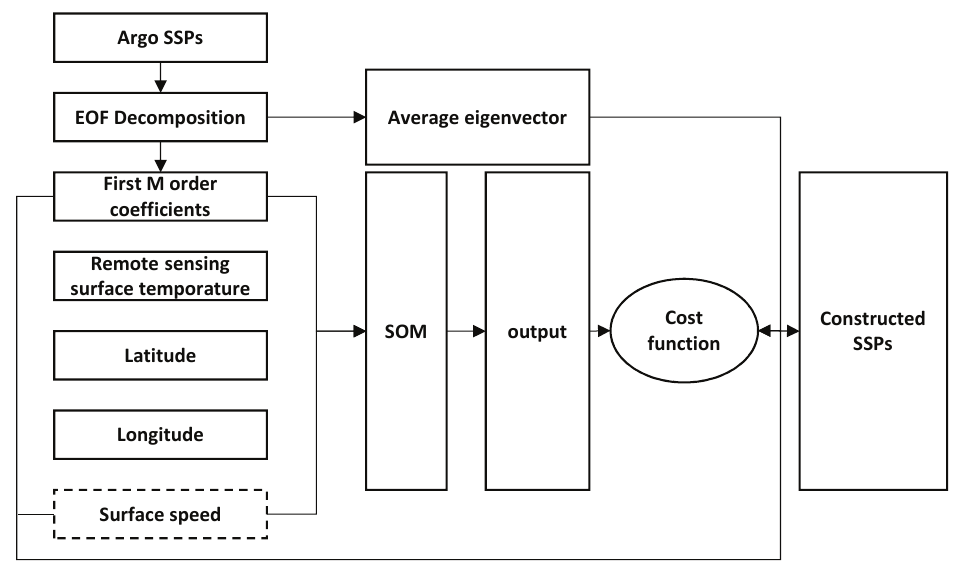}
	\caption{SOM model for SSP inversion.}
	\label{fig10}
\end{figure}

\indent In 2022, Li proposed a full--ocean--depth SSP inversion method based on self-organizing map (SOM) neural network \cite{Li2022Acoustic}. The model structure is shown in Fig.~\ref{fig10}. This method fully utilizes  empirical SSP data and ocean surface temperature data observed from the global satellite remote sensing system,  and optionally incorporates ocean surface sound speed measurement data, providing a feasible solution for constructing SSPs without sound field observation data. Compared to the SSP estimation method based on RBF neural networks, it can construct different spatial sound speed distributions to better reflect the spatial differences in sound speed distribution. But due to the model's reliance on ocean surface temperature and sound speed data, it is not possible to fully describe the distribution of sound speed in the seasonal thermocline. Therefore, there is still great space for improvement in the accuracy of constructing the full--ocean--depth SSP.

\subsubsection{Typical sound field measurement mode}

\begin{figure}[!t]
	\centering
	\subfloat[]{\includegraphics[width=2.5in]{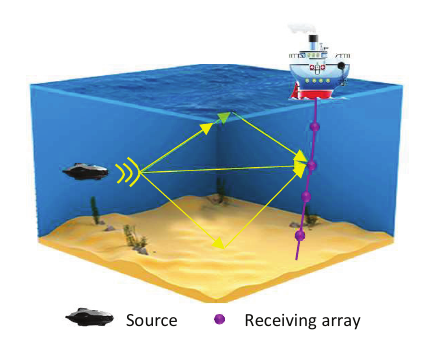}%
		\label{fig11a}}
	\hfil
	\subfloat[]{\includegraphics[width=2.5in]{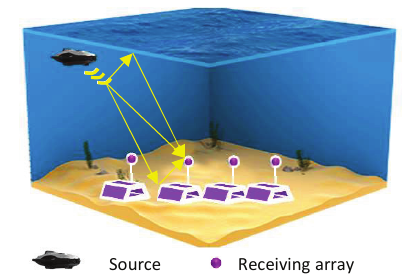}%
		\label{fig11b}}
	\caption{Sound field data measurement. (a) Vertical array. (b) Horizontal array.}
	\label{fig11}
\end{figure}
\indent In the framework of MFP, CS, and DL for SSP inversion, the accuracy of using sound field observation data for sound speed field inversion is still higher than that without sonar observation data at present. In terms of sound field information measurement, the Scripps Institute of Oceanography of the United States early adopted a single input single output (SISO) system for sound field measurement, using multipath signals to invert the environmental sound speed distribution \cite{cornuelle1989ocean}. In order to increase the information content of sound field data measurement and reduce the impact of environmental noise on measurement accuracy, the US Navy Laboratory used a single input multiple output (SIMO) system \cite{tolstoy1991acoustic} to validate the MFP sound speed inversion algorithm, which is implemented in the form of a ship towed vertical array as shown in Fig~\ref{fig11a}.

\indent The Institute of Acoustics, Chinese Academy of Sciences, has proposed a method of using seabed horizontal arrays for MFP to invert SSPs in response to the unstable attitude of vertical arrays, and successfully obtained shallow water sound speed inversion results \cite{li2010inversion,li2015sound}, as shown in Fig~\ref{fig11b}. However, due to the difficulty in recycling and reusing seabed arrays, the economic cost of equipment production is high.

\section{Comparison of methods for constructing underwater SSPs}
\begin{table*}[!htbp]
	\caption{comparizon of different methods for constructing underwater SSPs\label{table3}}
	\centering
	\begin{tabular}{|c|c|c|c|c|c|c|}
		\hline
		\multirow{2}*{Method} & \multirow{2}*{Sonar data}  & \multirow{2}*{Accuracy}  & \multicolumn{2}{c|}{Time consumption} & \multirow{2}*{Advantage} & \multirow{2}*{Disadvantage} \\
		\cline{4-5}
		\multirow{2}*{} & \multirow{2}*{} & \multirow{2}*{} & Preparation stage & Constructino stage & \multirow{2}*{} & \multirow{2}*{} \\
		\hline
		\makecell{CTD/SVP\\\cite{SBE2023,SonTek2023,OceanSeven2023,SeaSun2023,RBR2023,OST2023}} & -no & high & - & very long & accurate & \makecell{very long period;\\ high economic costs} \\
		\hline
		\makecell{XCTD\\ \cite{XCTD2023,Trampp2012Upper,Shi2015Expendable,Cheng2016SSP,Huang2023Fast}} & no & high & - & long & accurate; portable & \makecell{long period;\\limited depth coverage}\\
		\hline
		\makecell{EOF-MFP\\\cite{tolstoy1991acoustic,zhang2002beam,zhang2012inversion,zhang2013inversion,zheng2017improved,zhang2005thestudy,tang2006sound,sun2016inversion,zhang2015inversion}} & yes & good & very short & medium & near real--time & \makecell{complex optimization\\ object searching} \\
		\hline
		\makecell{Dictionary learning MFP \\\cite{Bianco2017Dictionary,Bianco2019Machine}}& yes & good & short & medium & near real--time & \makecell{complex optimization\\ object searching} \\
		\hline
		\makecell{CS \\\cite{Bianco2016CS,choo2018compressive}} & yes & medium & short & short & \makecell{sparse representation\\ low storage space;\\not bad real--time} & \makecell{accuracy loss\\by linear representation} \\
		\hline
		\makecell{ANN \\\cite{Stephan1995Tomographic,huang2018underwater}} & yes & good & long & very short & real--time & \makecell{complex preparation stage;\\ weak noise reistance;\\overfitting problem} \\
		\hline
		\makecell{AEFMNN \\\cite{Huang2019Collaborating}} & yes & good & long & very short & \makecell{real--time;\\good robustness} & \makecell{complex preparation stage;\\overfitting problem} \\
		\hline
		\makecell{TDML \\\cite{Huang2023Meta}} & yes & good & very long & very short & \makecell{real--time;\\anti--overfitting} & complex preparation stage \\
		\hline
		\makecell{RBF \\\cite{Yu2020SSP,Wang2022Ocean}} & no & low & long &very short  &  \makecell{real--time;\\no sonar data}  &  \makecell{weak time and space\\resolution ability} \\
		\hline
		\makecell{SOM \\\cite{Li2022Acoustic}} & no & low & long &very short  &  \makecell{real--time;\\no sonar data}  &  \makecell{insufficient accuracy\\in the deep--ocean part}\\
		\hline
	\end{tabular}
\end{table*}
\indent Table~\ref{table3} summarizes the basic attributes of different methods for constructing SSPs. The construction accuracy is divided into high, good, medium, and low levels from the best to the worst. The time cost is divided into the preparation phase and the inversion phase, and the preparation phase can be completed offline. Therefore, the time cost performance of the inversion phase is more critical, and is divided into 5 levels from the best to the worst: very short, short, medium, long, and very long. Due to the fact that CTD, SVP, XCTD and other devices belong to direct measurement methods, their time cost is calculated in hours and minutes, while MFP, CS, and DL models belong to sound speed inversion methods, and their time cost is usually calculated in minutes and seconds. Therefore, the "very long, long" levels of time cost is much greater than the "medium, short, and very short" levels of time cost.

\indent Through comparison, it can be seen that the biggest advantage of the direct measurement method lies in its accurate measurement results, which can usually be considered as true values. The main disadvantage is that the measurement period is long and it is difficult to meet the real-time requirements for estimating the sound speed distribution in underwater applications. In the SSP inversion methods, the construction of SSPs using sound field observation data has higher accuracy than the case without sound field observation data. Compared to MFP, CS improves real--time performance but sacrifices accuracy. Compared to the two methods mentioned above, the neural network model performs better in accuracy and real-time performance, but in essence, its real-time advantage is at the cost of sacrificing the time cost of the preparation stage. The biggest advantage of RBF and SOM models is that they are not constrained by sound field observation data, making them widely applicable to various regions around the world and greatly reducing the cost of underwater acoustic equipment. However, after training with designated spatio--temporal data, the generalization ability of RBF models in time and space resolution is insufficient, while the fitting accuracy of SOM models for seasonal thermocline and deep-sea isothermal layer is insufficient.

\indent With the development of sensing technology, robotics technology, and the long--term accumulation of empirical observation data, the overall trend of SSP construction methods is towards intelligence, real--time, and precision. On the one hand, the increase in data sources means that the selectivity of methods for constructing sound speed fields is improved. In different regions, elastic construction of sound speed fields based on the available data sources will be an excellent solution to meet the accuracy or real-time requirements of underwater users for estimating sound speed distribution in the future. On the other hand, in the absence of sound field observation data, how to use historical observation data and prior information that can be quickly measured through remote sensing or sea surface to intelligently and accurately construct a full ocean depth sound speed field is still a research point worth further exploration.

\section{Conclusion}
\indent This paper introduces and investigates the current research status of underwater sound speed field construction, which can be divided into two categories: direct measurement methods and sound speed inversion methods. For the direct measurement method, this paper investigates and lists the current mainstream commercial equipment and instruments, and lists their main performance parameters. For the sound speed inversion method, this paper summarizes the three major frameworks of MFP, CS, and DL based on the principles for constructing sound speed fields. By analyzing the principles of specific methods within each framework, the advantages and disadvantages of different sound speed field construction methods are compared, and the evolution trend of sound speed field construction methods is summarized, which is towards intelligent, real--time, and accurate development. The existing real-time construction methods for sound speed fields do not match the demand for sound field observation data with the production cost of expensive underwater sound field measurement equipment, which limits the universality of underwater sound speed inversion methods. Therefore, how to use limited prior information for intelligent and high--precision construction of full--ocean--depth sound speed fields in the absence of sound field observation data is a major trend in future research, and how to use multi-source observation data to provide elastic estimation services for underwater users with different accuracy and real--time requirements of sound speed distribution is another important trend in future research. The work of this paper will provide reference for subsequent research on the construction of underwater sound speed fields.


\section*{Acknowledgments}
This work was supported by Natural Science Foundation of Shandong Province (ZR2023QF128), China Postdoctoral Science Foundation (2022M722990), Qingdao Postdoctoral Science Foundation (QDBSH20220202061), National Natural Science Foundation of China (62271459), National Defense Science and Technology Innovation Special Zone Project: Marine Science and Technology Collaborative Innovation Center (22-05-CXZX-04-01-02), Fundamental Research Funds for the Central Universities, Ocean University of China (202313036).

\bibliographystyle{IEEEtran}
\bibliography{IEEEabrv,draft_survey}
\vspace{-10mm}
\begin{IEEEbiography}[{\includegraphics[width=1in,height=1.25in,clip,keepaspectratio]{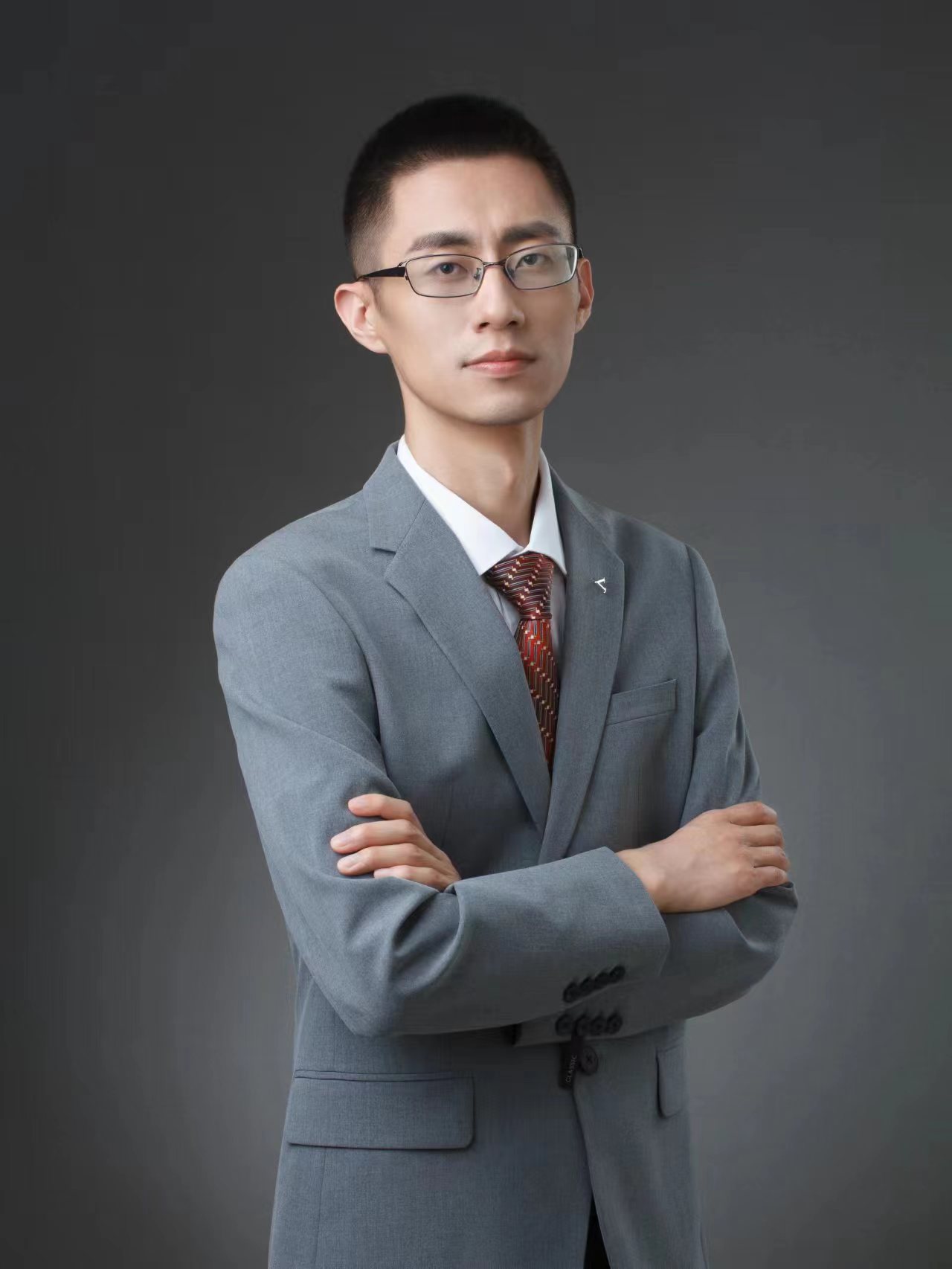}}]{Wei Huang} (IEEE S'18-M'22) received the Ph.D. degree in communication and information system with the School of Electronic Information from Wuhan University, China in 2021. He has published more than 10 SCI/EI research papers.He is now a lecturer and postdoctor at the Faculty of Information Science and Engineering, Ocean University of China.

His current research interests include underwater acoustic tomography, underwater acoustic communication and localization system, and underwater intelligent data processing.
\end{IEEEbiography}
\vspace{-10mm}
\begin{IEEEbiography}[{\includegraphics[width=1in,height=1.25in,clip,keepaspectratio]{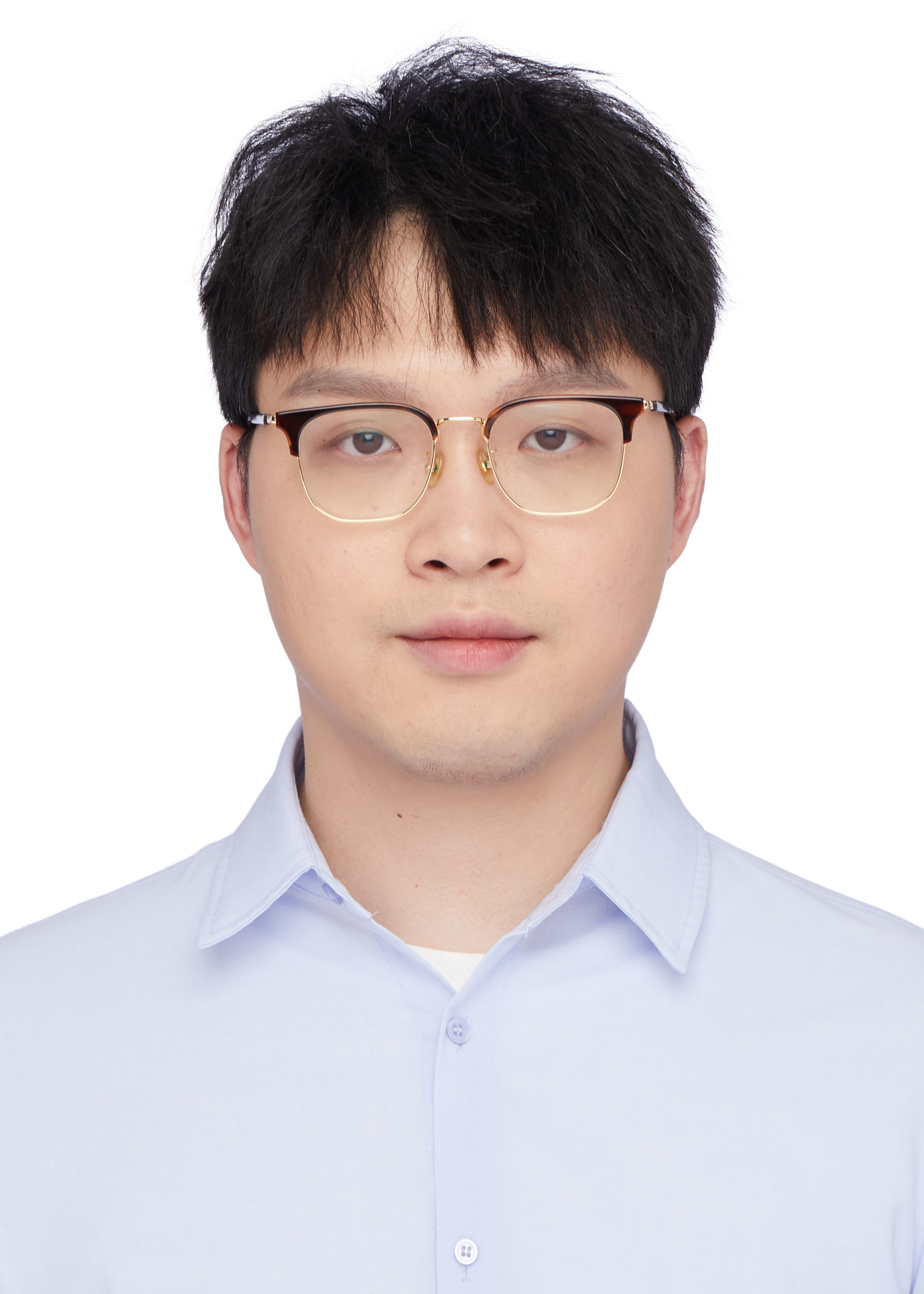}}]{Jixuan Zhou} received the B.Eng. degree in software engineering from the School of Software, Central South University, Changsha, China, in 2012, and the M.Eng. degree in School of Electronic Information, Wuhan University, Wuhan, China, in 2015. He received the Ph.D. degree in Internet of Things, Mobile Crowd Sensing with the School of Electronic Information, Wuhan University, Wuhan, China. His current research interests include participatory sensing, urban sensing, and underwater internet of things. 
\end{IEEEbiography}
\vspace{-10mm}
\begin{IEEEbiography}[{\includegraphics[width=1in,height=1.25in,clip,keepaspectratio]{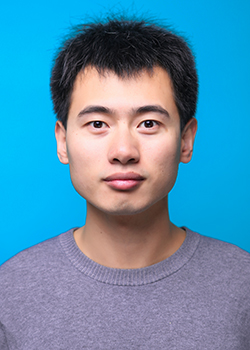}}]{Fan Gao} is now an associate professor with the School of Space Science and Physics, Shandong University. He received his Ph.D. degree in Geodesy and Surveying Engineering from the University of Chinese Academic of Sciences in 2016. His research interests include GNSS-R altimetry, GNSS software-defined receiver, satellite constellation design, precise orbit determination, and underwater navigation.
\end{IEEEbiography}
\vspace{-5mm}
\begin{IEEEbiography}[{\includegraphics[width=1in,height=1.25in,clip,keepaspectratio]{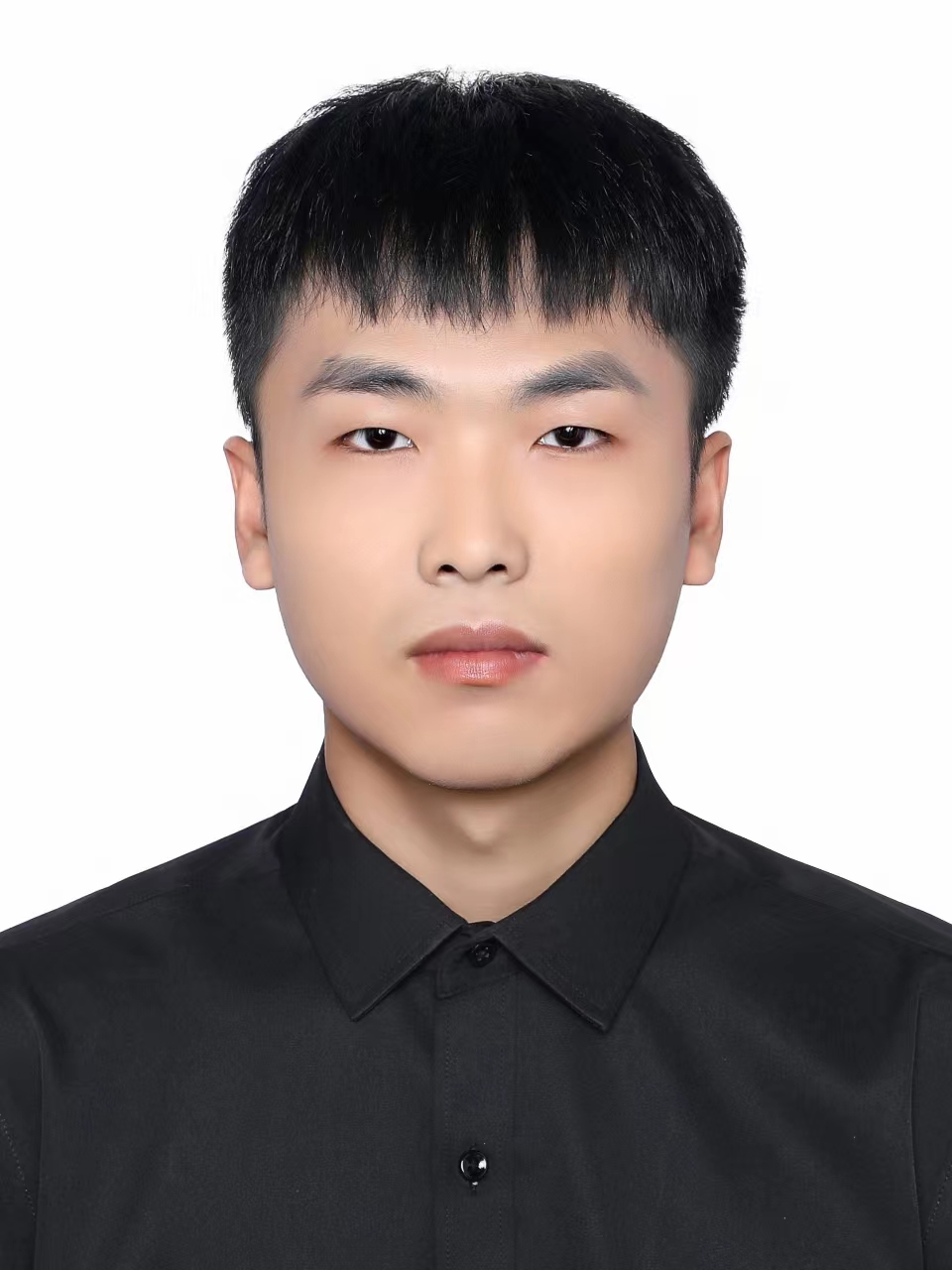}}]{Jiajun Lu} received the the bachelor's degree in electronic information science and technology from Heze University, China. He is now persuing the master's degree in Commmunication Engineering in Ocean University of China. His current research interest focuses on underwater sound speed prediction.
\end{IEEEbiography}
\vspace{-5mm}
\begin{IEEEbiography}[{\includegraphics[width=1in,height=1.25in,clip,keepaspectratio]{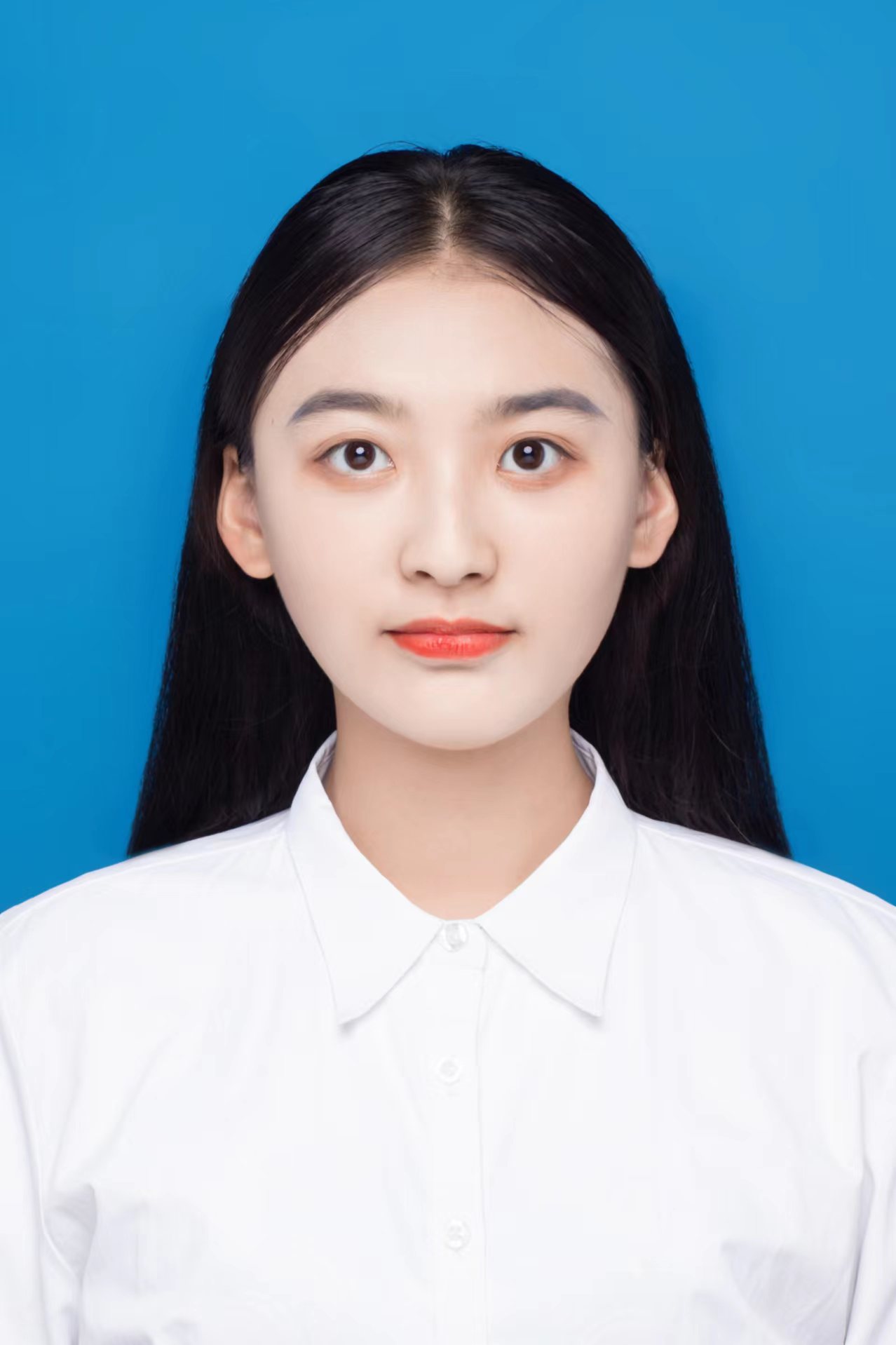}}]{Sijia Li} received the the bachelor's degree in communication engineering from Shandong Normal University, China. She is now persuing the master's degree in Commmunication Engineering in Ocean University of China. Her current research interest focuses on underwater sound speed construction.
\end{IEEEbiography}

\begin{IEEEbiography}[{\includegraphics[width=1in,height=1.25in,clip,keepaspectratio]{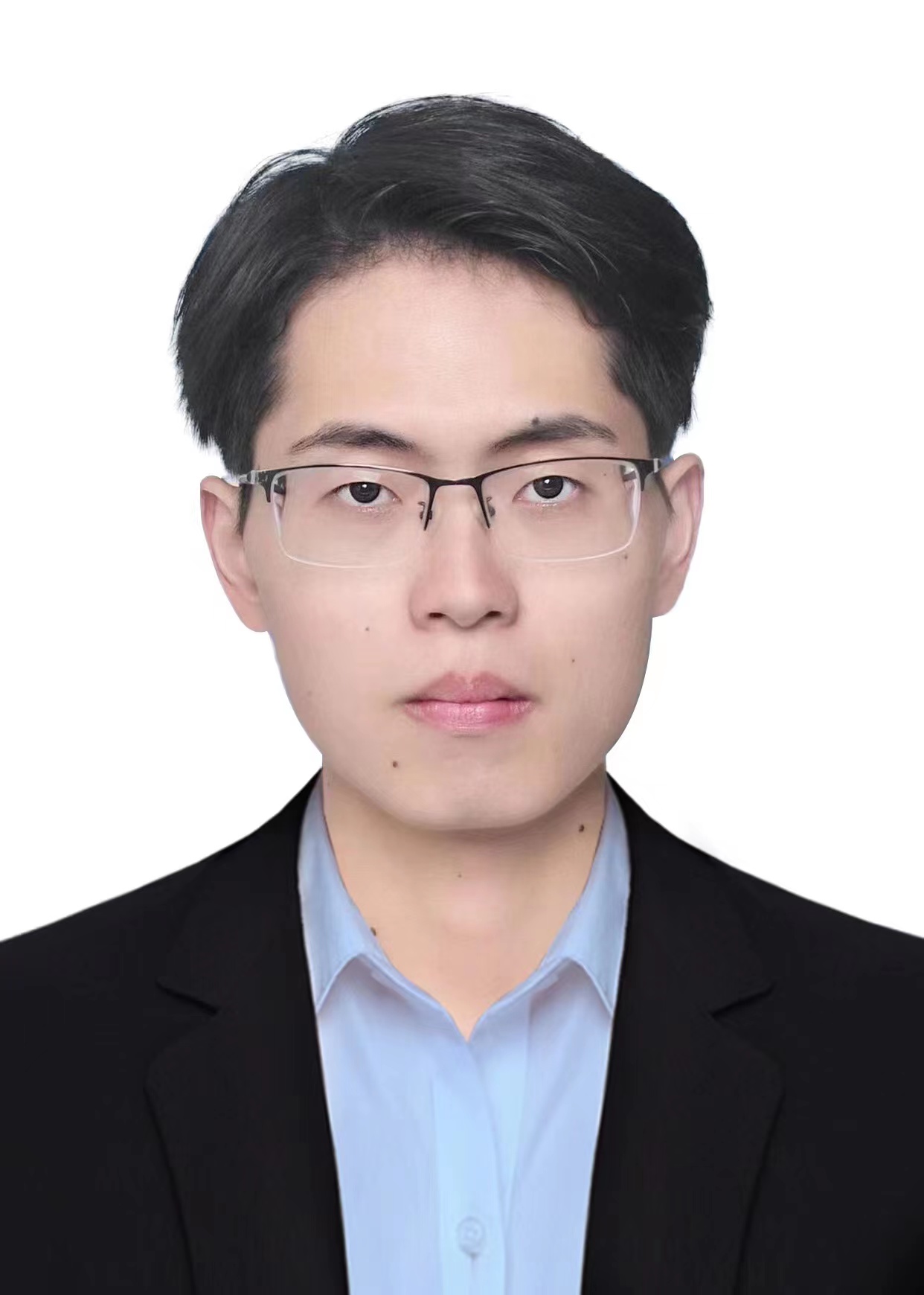}}]{Pengfei Wu} received the the bachelor's degree in communication engineering from Tibet University, China. He is now persuing the master's degree in Commmunication Engineering in Ocean University of China. His current research interest focuses on underwater sound speed construction.
\end{IEEEbiography}

\begin{IEEEbiography}[{\includegraphics[width=1in,height=1.25in,clip,keepaspectratio]{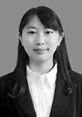}}]{Junting Wang} obtained her BSc and MSc degrees from Shandong Agricultural University and China University of Petroleum (East China) in 2014 and 2017, respectively. She obtained her Ph.D. degree from Shandong University in 2022. Her research interests include underwater acoustic positioning, multi-sensor integrated navigation positioning and the inversion of sound speed field, etc.
\end{IEEEbiography}

\begin{IEEEbiography}[{\includegraphics[width=1in,height=1.25in,clip,keepaspectratio]{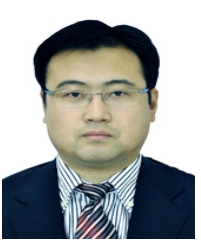}}]{Hao Zhang} (Senior Member, IEEE) received the bachelor's degree in communication engineering from Shanghai Jiao Tong University in 1994, and received the Ph.D. degree in Electronic Engineering from University of Victoria in 2004. He is now a professor and the director of the institute of Ocean Communication at Ocean University of China, and a visting professor at the University of Victoria, Canada. He is one of the Leading Talents in the National "Ten Thousand Talents Plan", he is also one of the Outstanding Talents in the New Century by the Ministry of Education, and a winner of Outstanding Young Scholars in Natural Science in Shandong Province. 
	
He has published more than 150 SCI/EI research papers. His current interests include underwater signal processing, wireless communication, navigation and communication of satellite system, underwater sensor networks.
\end{IEEEbiography}

\begin{IEEEbiography}[{\includegraphics[width=1in,height=1.25in,clip,keepaspectratio]{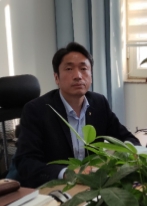}}]{Tianhe Xu} received his Ph.D. and M.S. degrees in Geodesy from the Zhengzhou Institute of Surveying and Mapping of China in 2004 and 2001. He is now a professor at the Institute of Space Sciences in Shandong University, Weihai. He is a chief scientist of National Key Research and Development Program of China. He has published more than 150 SCI/EI papers with citation of more than 3000 times. He has received 2 provincial and ministerial level special awards for scientific and technological progress, 3 first prizes, 7 second prizes, and 1 third prize.	His research interests include satellite navigation, orbit determination, satellite gravity data processing and quality control.
\end{IEEEbiography}

\vfill

\end{document}